\documentclass[aps,prb,showpacs,twocolumn]{revtex4}
\usepackage{amsfonts}
\usepackage{amssymb}
\usepackage{amsmath}
\usepackage{graphicx}
\usepackage{verbatim}
\usepackage{color}

\begin{document}

\title{Quantum theory of the charge stability diagram of semiconductor double quantum dot systems}
\author{Xin Wang, Shuo Yang, and S. Das Sarma}
\affiliation{Condensed Matter Theory Center, Department of Physics,
University of Maryland, College Park, MD 20742}

\begin{abstract}
We complete our recently introduced theoretical framework treating the double quantum dot system with a generalized form of Hubbard model. The effects of all quantum parameters involved in our model on the charge stability diagram are discussed in detail. A general formulation of the microscopic theory is presented, and truncating at one orbital per site, we study the implication of different choices of the model confinement potential on the Hubbard parameters as well as the charge stability diagram. We calculate the charge stability diagram keeping three orbitals per site and find that the effect of additional higher-lying orbitals on the subspace with lowest-energy orbitals only can be regarded as a small renormalization of Hubbard parameters, thereby justifying our practice of keeping only the lowest-orbital in all other calculations. The role of the harmonic oscillator frequency in the  implementation of the Gaussian model potential is discussed, and the effect of an external magnetic field is identified to be similar to choosing a more localized electron wave function in microscopic calculations. The full matrix form of the Hamiltonian including all possible exchange terms, and several peculiar charge stability diagrams due to unphysical parameters are presented in the appendix, thus emphasizing the critical importance of a reliable microscopic model in obtaining the system parameters defining the Hamiltonian.
\end{abstract}

\begin{comment}
73.21.-b	Electron states and collective excitations in multilayers, quantum wells, mesoscopic, and nanoscale systems (for electron states in nanoscale materials, see 73.22.-f)
73.21.La	Quantum dots
03.67.Lx	Quantum computation architectures and implementations
73.63.-b	Electronic transport in nanoscale materials and structures (see also 73.23.-b Electronic transport in mesoscopic systems)
73.63.Kv	Quantum dots
85.30.-z	Semiconductor devices (for photodiodes, phototransistors, and photoresistors, see 85.60.Dw; for laser diodes, see 42.55.Px; for semiconductor-based solar cells, see 88.40.-j; for applications of dielectric films in silicon electronics, see 77.55.df)
71.10.-w	Theories and models of many-electron systems
85.35.Gv	Single electron devices
\end{comment}

\maketitle

\section{Introduction}

The idea of quantum computation, introduced two decades ago,\cite{Feynman.86} has attracted intense research interest because of its ability to provide novel solutions to certain problems that have been deemed intractable on a classical computer.\cite{Shor.94} The fundamental unit of a quantum computer is the qubit, which is implemented using quantum two-level (and sometimes multi-level) systems. Many physical systems have been put forward as candidates on which qubits can be embedded, such as nuclear magnetic resonance,\cite{Vandersypen.00} cavity QED,\cite{Sleator.95} and trapped ions.\cite{Cirac.95} Among these proposals, the solid state systems based on semiconductor quantum dots stand out as most promising for the implementation of qubits,\cite{Loss.98,DiVincenzo.00,Levy.02,Koppens.06} since modern semiconductor industry allows great scalability for these systems.\cite{Kane.98,Vrijen.00,Friesen.03,Taylor.05} Long coherence times have been experimentally achieved in both GaAs\cite{Kikkawa.98,Amasha.08,Koppens.08,Barthel.10z,Bluhm.11} and Si\cite{Tyryshkin.03,Morello.10,Simmons.10,Xiao.10a} semiconductor devices. One-qubit and two-qubit manipulations have been demonstrated in the GaAs quantum dot systems.\cite{Petta.05,Koppens.06,Foletti.09,Laird.10,vanWeperen.11}
In silicon quantum dot systems, despite difficulties such as the fabrication-induced disorder\cite{Nordberg.09} and the valley degree of freedom,\cite{dassarma.05,Culcer.10, Goswami.07,Borselli.11} high tunability has been reported\cite{Simmons.09,Lai.10} as a milestone toward coherent control of the qubit. 

Two main effects are invariably present in all these experiments involving semiconductor quantum dots: the classical Coulomb interaction and the quantum fluctuations. On one hand, due to Coulomb blockade,\cite{Livermore.96} the electron configurations in the quantum dot system are precisely controlled by electrostatic potentials (gate voltages). The details of electron configurations can be extracted by an adjacent quantum point contact,\cite{Elzerman.03} and visualized as the charge stability diagram.\cite{Hofmann.95,DiCarlo.04} On the other hand, the quantum effect, in particular the exchange interaction,\cite{Scarola.04,LiQZ.10} plays an essential role in the qubit manipulation. The exchange interaction constitutes two-qubit operations in the Loss-DiVincenzo proposal\cite{Loss.98} (which uses single-electron spin-up/down states as the qubit). In the double dot singlet/triplet proposal, the exchange interaction controls the singlet-triplet level splitting thus, along with an inhomogeneous magnetic field, provides arbitrary rotation along the Bloch sphere and thereby achieves full single-qubit control.\cite{Levy.02,Petta.05,Taylor.05,Taylor.07} In a three-dot scheme,\cite{Laird.10} the need for an inhomogeneous magnetic field is eliminated, and the exchange interaction alone suffices for universal quantum computation.

The charge stability diagram is fundamental for spin-based quantum computation: it is the starting point for all subsequent qubit manipulation\cite{Petta.08,Reilly.08,Reilly.10} and information readout.\cite{Barthel.09,Barthel.10a} Previous studies of the charge stability diagram have focused on the classical Coulomb effects using primarily the capacitance circuit model,\cite{Beenakker.91,Wiel.03,Hanson.07,Schroer.07} which parametrizes the on-site and inter-site Coulomb interaction as capacitances. The charge stability diagram is found through the electrostatic energy for a given electron configuration on the double dot system. Although this classical theory has satisfactorily explained many aspects of experiments (especially the cases where the quantum fluctuations are weak), there are cases where the quantum effects, inevitably intertwining with the Coulomb interaction, lead to (sometimes substantial) deformation of the charge stability diagram from that predicted by the capacitance circuit model.\cite{Simmons.09,Lai.10,dassarma.Si,Yang.11} A theory capable of reconciling both the classical and quantum effects becomes necessary, which will not only help us understand the quantum aspect of the charge stability diagram, but will also substantially extend its usefulness. 

Attempts to explain the quantum aspects of the charge stability diagram have been many. Most of them have employed the quantum-mechanical two-level model to study the tunneling of a single electron from one dot to the other.\cite{DiCarlo.04,Petta.04,Hatano.05,Huttel.05,Pioro.05} From the probability crossover of the two eigenstates (sometimes noted as the ``excess charge'') one can reliably extract the parameter for quantum-mechanical electron hopping (or tunnel coupling) $t$. A somewhat similar work\cite{ZhangLX.06} has studied the influence of the microscopically calculated exchange interaction on the charge stability diagram. On the other hand, interest in applying Hubbard-like strongly correlated models to the quantum dot system\cite{Stafford.94,Kotlyar.98} arises in contexts related to the collective Coulomb blockade. In Ref.~\onlinecite{Jefferson.96}, a Hamiltonian whose form is essentially very similar to our proposed generalized Hubbard model\cite{Yang.11} has been derived using the so-called ``pocket-state'' method. However, the power of Hubbard-like models to describe the charge stability diagram has long been neglected until Gaudreau and collaborators\cite{Gaudreau.06,Korkusinski.07}  have applied the Hubbard model to triple, but not double, quantum dot system.

In a previous publication,\cite{Yang.11} we have introduced a generalized Hubbard model as a quantum generalization of the capacitance circuit model. We have shown that, with all quantum fluctuations suppressed, the generalized Hubbard model becomes the extended Hubbard model, which can be mapped to the classical capacitance circuit model exactly.  The main advantage of our approach is that all terms allowed by symmetry arguments are kept in the Hamiltonian, which naturally accommodates all possible kinds of quantum fluctuations. We have further recognized that these quantum effects cannot be included in the capacitance model in any \emph{ad hoc} manner. Because the electron occupancies on each dot are no longer good quantum numbers, it breaks the basic assumption of the capacitance model that the electrostatic energy of the system is expressed in terms of electron occupancies on each dot, which are assumed to be integer numbers. Therefore, our generalized Hubbard model can be viewed as the quantum generalization of the classical capacitance circuit model, with the individual electron occupancy on each dot being a fluctuating quantum variable instead of being a fixed number as in the capacitance model.

The quantum fluctuations may perturb the charge stability diagram in a substantial way.
In Ref.~\onlinecite{Yang.11} we discussed in particular the rounding effect of the boundary lines due to the electron hopping (tunnel coupling) $t$. A microscopic calculation was outlined to control the parameters in the generalized Hubbard model, and we have shown how the charge stability diagram changes with the variation of quantum fluctuations induced by the deformation of the microscopic confinement potential. In Ref.~\onlinecite{dassarma.Si} we applied our theory to quantitatively explain two set of experiments\cite{Simmons.09,Lai.10} on the silicon system and calculated, in particular,  the tunnel coupling as a function of the height of the central potential barrier. The results are found to well describe the experimental data after appropriately choosing model parameters.  

In this paper, we address problems that have not been fully explored in our previous publications.\cite{dassarma.Si, Yang.11} The primary goal of this paper is to provide a comprehensive picture of the theoretical framework that we have introduced in Refs.~\onlinecite{Yang.11, dassarma.Si}, thereby completing this series of study on the charge stability diagram of the double quantum dot system. In Refs.~\onlinecite{Yang.11, dassarma.Si}, we have proposed that the quantum fluctuations are essential for a in-depth understanding of the charge stability diagram, and have focused our attention on the tunnel coupling as it is manifestly the most significant quantum parameter. However, other parameters such as the spin-exchange, pair-hopping and the occupation-modulated hopping are present as well, albeit with smaller amplitudes. One of the central aims of this paper is to quantitatively estimate the effect of all these quantum parameters, which define the full quantum Hubbard model, on the charge stability diagram. Moreover, although we have outlined the microscopic theory and performed calculation in some specific cases in Refs.~\onlinecite{Yang.11, dassarma.Si}, there are several problems that have not been sufficiently covered in the previous publications. First, Ref.~\onlinecite{Yang.11} has focused on the biquadratic model potential and, since this model is rather special and involves only a few parameters, a dimensionless quantity $\eta$ combining the height of central potential barrier and the energy of electron states is identified to quantify the quantum fluctuations and completely determine the geometry of the charge stability diagram. In Ref.~\onlinecite{dassarma.Si} the results calculated from both the biquadratic model and the Gaussian model confinement potentials have been compared to experiments side-by-side. However a direct comparison of the two models in terms of the Hubbard parameters is missing. Although experimentally the electrostatic potentials dominate quantitatively, the detailed form of the actual confinement potential is obviously unknown. Thus it is important to understand the consequences of different choices of model potentials, under similar electrostatic situations, on the Hubbard parameters as well as the charge stability diagram. We will therefore compare two models, biquadratic and Gaussian, in this paper with respect to the charge stability diagram. Second, all calculations presented in our previous publications\cite{dassarma.Si, Yang.11} as well as most of this paper are done assuming that each dot contains a single orbital and is only allowed to hold up to two electrons. We will discuss the case where this constraint is lifted to allow three orbitals per site, and projecting to the single-orbital subspace we study the effect of additional higher-lying orbitals on this single-orbital subspace.
Last, we will also discuss the role of the microscopic harmonic oscillator frequency in the Gaussian model potential, as well as the effect of external magnetic fields. All of these are new and not discussed in our previous publications.\cite{dassarma.Si, Yang.11}

The remainder of this paper is organized as follows. In Sec.~\ref{sec:microgen} we present a general microscopic formulation of the problem. In Sec.~\ref{sec:Hubgen} we discuss in detail the generalized Hubbard model proposed in Ref.~\onlinecite{Yang.11}. In Sec.~\ref{sec:Hubres} we show charge stability diagrams calculated directly from the generalized Hubbard model and discuss the effect of various parameters on the stability diagram. In Sec.~\ref{sec:microres} we present a very detailed discussion on the microscopic calculation, including the effects of different choices of confinement potential, the additional higher-lying orbitals, the role of the harmonic oscillator frequency in the Gaussian model potential and the effect of the magnetic field. Sec.~\ref{conclusion} is the conclusion. The full matrix form of the Hamiltonian and several cases with extreme (and possibly unrealistic) parameters are shown in the appendices. 

\section{Microscopic theory: General formulation}\label{sec:microgen}

We start with the general Hamiltonian of a system with $N$ electrons, which
consists of three parts:
\begin{equation}
H\left(N\right) = H_{0}+H_{C}+H_{Z}.\label{firstH}
\end{equation}
$H_{0}$ is the sum of the single-particle
Hamiltonian for each electron, $H_{0}=\sum_{i=1}^{N}h\left(\boldsymbol{r}_{i}\right)$.
$H_{C}$ is the Coulomb interaction between electrons,
\begin{equation}
H_{C} = \sum_{i=1}^{N-1}\sum_{i'=i+1}^{N}\frac{ke^{2}}{
\left|\boldsymbol{r}_{i}-\boldsymbol{r}_{i'}\right|},
\end{equation}
where $k=1/(4\pi\varepsilon_0\varepsilon)$.
$H_{Z}$ is the Zeeman energy,
\begin{align}
H_{Z} = g^{*}\mu_{B}B\sum_{i=1}^{N}S_{i}^{z}=E_{B}S^{z},
\end{align}
where $E_{B}=g^{*}\mu_{B}B$, $S^{z}=\sum_{i=1}^{N}S_{i}^{z}$.
We neglect the spin-orbit couplings for simplicity. We further note that in our work, we do not include the interaction between the electrons on the quantum dots and the environment. Therefore Eq.~\eqref{firstH} is understood as describing the electrons on the quantum dots only. In fact, many environmental factors contribute to the decoherence process of the quantum dot system, such as the hyperfine interaction with the nuclear spin bath,\cite{Witzel.06,Cywinski.09,Taylor.07,Reilly.08,Reilly.10} the coupling to background impurities\cite{Gimenez.09,Nguyen.11} and the phonon modes.\cite{Storcz.05,Stavrou.05,Stano.06,Harbusch.10,Hu.11} In this work, we concentrate on the electronic interaction in the quantum dot system and highlight its consequence on the charge stability diagram. Therefore at this early stage of the theory we completely neglect the coupling to the environment. The elucidation of this problem is important in future studies.

The single-particle Hamiltonian $h\left(\boldsymbol{r}\right)$ can be written as
\begin{align}
h\left(\boldsymbol{r}\right) = \frac{1}{2m^{*}}\left[\boldsymbol{p}-e\boldsymbol{A}\left(\boldsymbol{r}\right)\right]^{2}+V\left(\boldsymbol{r}\right),
\end{align}
where $m^{*}$ is the effective mass of the electrons and $V(\boldsymbol{r})$ is the potential (typically double-welled in the case of the triplet/singlet spin qubit) confining electrons on the $xy$-plane.
We allow for a magnetic field $\boldsymbol{B}$ applied along the $z$ axis, which
couples to the electrons via the vector potential $\boldsymbol{A}$.
Here we have assumed that all electrons are experiencing the same
magnetic field (in actual experiments this may not be the case). 

There are different choices of the detailed form of the confinement potential, but we assume that it is approximately parabolic  around the minima of the potential well so that the ground state single-particle
wave functions are harmonic oscillator states. In this case, the single-particle Hamiltonian $h\left(\boldsymbol{r}\right)$
can be written as the sum of the Fock-Darwin Hamiltonian $H_{\mathrm{FD},j}$
and some non-harmonic potential $W_j\left(\boldsymbol{r}\right)$, $h\left(\boldsymbol{r}\right)=H_{\mathrm{FD},j}+W_{j}\left(\boldsymbol{r}\right)$,
with
\begin{align}
H_{\mathrm{FD},j} &= \frac{1}{2m^{*}}\left[\boldsymbol{p}-e\boldsymbol{A}\left(\boldsymbol{r}\right)\right]^{2}+\frac{1}{2}m^{*}\omega_{j}^{2}\left(\boldsymbol{r}-\boldsymbol{r}_{j}\right)^{2},  \\
W_{j}\left(\boldsymbol{r}\right) &= V\left(\boldsymbol{r}\right)-\frac{1}{2}m^{*}\omega_{j}^{2}\left(\boldsymbol{r}-\boldsymbol{r}_{j}\right)^{2}.
\end{align}
Here, $\omega_j$ is the corresponding harmonic oscillator frequency for dot $j$. We denote $\boldsymbol{r}_j'=\boldsymbol{r}-\boldsymbol{r}_{j}$ and express $\boldsymbol{r}_j'$ as $({r}_j'\cos\theta_j',{r}_j'\sin\theta_j')$ in Cartesian coordinates. 
The eigenfunctions of the Fock-Darwin Hamiltonian at the $j$th dot
are the Fock-Darwin states satisfying
\begin{align}
H_{\mathrm{FD},j}\left|\varphi_{j,n_{j},m_{j}}\left(r_{j}',\theta_{j}'\right)\right\rangle  &= \epsilon_{n_{j+},n_{j-}}\left|\varphi_{j,n_{j},m_{j}}\left(r_{j}',\theta_{j}'\right)\right\rangle .
\end{align}
where 
\begin{align}
\epsilon_{n_{j+},n_{j-}}=\hbar\omega&_{j+}\left(n_{j+}+1/2\right)+\hbar\omega_{j-}\left(n_{j-}+1/2\right),\\
\varphi_{j,n_{j},m_{j}}\left(r_{j}',\theta_{j}'\right)&=\exp\left[-i\left(eB/2\hbar\right)\left(x_{j}y-y_{j}x\right)\right]\notag\\
&\quad\cdot\psi_{m_{j}}\left(\theta_{j}'\right)R_{n_{j},m_{j}}\left(r_{j}'\right),
\end{align}
with
\begin{align}
\psi_{m_{j}}\left(\theta_{j}'\right)=e^{im_{j}\theta_{j}'}/\sqrt{2\pi}
\end{align}
and
\begin{align}
R_{n_j,m_j}\left(r_j'\right)&=\sqrt{2n^r_j!/\left(n^r_j+\left|m_{j}\right|\right)!}/l_{j0}\left(r_{j}'/l_{j0}\right)^{\left|m_{j}\right|}\notag\\
&\quad\cdot\exp\left[-r_{j}'^{2}/2l_{j0}^{2}\right]\mathcal{L}_{n^r_j}^{\left|m_{j}\right|}\left(\frac{r_{j}'^{2}}{l_{j0}^{2}}\right).
\end{align}

The index $n_j=0,1,\cdots$ is the principal quantum number, $m_{j}=-n_j,-n_j+2,
\cdots,n_j-2,n_j$ is the azimuthal quantum number, and $n^r_j=\left(n_j-\left|m_j\right|\right)/2$
is the radial quantum number. The pairs of quantum numbers $\left(n_{j},m_{j}\right)$
and $\left(n_{j+},n_{j-}\right)$ are related by $n_{j}=n_{j-}+n_{j+}$,
$m_{j}=n_{j-}-n_{j+}$. $\mathcal{L}_{n^r_j}^{\left|m_{j}\right|}$
denotes the Laguerre polynomials, $l_{j0}=\sqrt{\hbar/eB}{\big /}\sqrt[4]{1/4+\omega_{j}^{2}/\omega_{c}^{2}}$, and
\begin{equation}
\omega_{j\pm}=\sqrt{\omega_{j}^{2}+\omega_{c}^{2}/4}\pm\omega_{c}/2,\label{eqfrequency}
\end{equation}
where the Larmor frequency $\omega_{c}=eB/m^{*}$.

These Fock-Darwin states can be used either  to solve the Hamiltonian approximately, or to generate an equivalent tight-binding model. In Ref.~\onlinecite{Yang.11}, we have obtained a generalized Hubbard model using the Fock-Darwin states. 
There are infinitely many Fock-Darwin states, and it is impossible
and unnecessary to keep all of them. For the $M$-dot system, we truncate the Fock-Darwin basis in each dot to
$\nu$ levels, i.e., $\left|\varphi_{l}\right\rangle =\left|\varphi_{j,n}\right\rangle$, where $j$ labels quantum
dots ($1\leq j\leq M$)  and $n$ denotes the number of energy levels that we have kept ($1\leq n\leq\nu$). We notice that
the Fock-Darwin states within one dot are orthogonal to each other, $\left\langle \varphi_{j,n}\right|\left.\varphi_{j,n'}\right\rangle =0$, but those from different dots are in general not orthogonal:
$\left\langle \varphi_{j,n}\right|\left.\varphi_{j,n'}\right\rangle \neq0$.
We therefore build a new set of orthogonal
bases by making the transformation\cite{Artacho.91}
\begin{align}
&\left(\left|\Psi_{1}\right\rangle~\left|\Psi_{2}\right\rangle~\cdots~\left|\Psi_{M\nu}\right\rangle\right)^{\mathrm{T}}
\notag\\
&={\bf O}^{-1/2}\left(
\left|\varphi_{1}\right\rangle~\left|\varphi_{2}\right\rangle~\cdots~\left|\varphi_{M\nu}\right\rangle\right)^{\mathrm{T}}.
\end{align}
Here ${\bf O}$ is the overlap matrix ($O_{l,l'}=\left\langle \varphi_{l}\right|\left.\varphi_{l'}\right\rangle $) generated by Fock-Darwin states in
the single particle subspace.
The new basis $\left|\Psi_{l}\right\rangle$ actually corresponds to, in a second-quantized form,
$c_{l,\sigma}^\dagger\left|0\right\rangle$ where $c_{l,\sigma}^\dagger$ creates an electron on site/orbital $l$ with spin $\sigma$. After the orthogonalization,
the second quantized form of $H_0$ and $H_C$ can be built, in a standard
way, as
\begin{align}
H_{0}' &= \sum_{l_1,l_2,\sigma}F_{l_1,l_2}c_{l_1,\sigma}^{\dagger}c_{l_2,\sigma}, \label{eqhamF1}\\
H_{C}' &= \sum_{l_1,l_2,l_3,l_4}\Bigl[G_{l_1,l_2,l_3,l_4}^{(1)}c_{l_1,\uparrow}^\dagger c_{l_2,\downarrow}^\dagger c_{l_3,\uparrow}c_{l_4,\downarrow}\notag\\
&+\sum_\sigma G_{l_1,l_2,l_3,l_4}^{(2)}c_{l_1,\sigma}^{\dagger}c_{l_2,\sigma}^{\dagger}c_{l_3,\sigma}c_{l_4,\sigma}\Bigr].
\end{align}
The coupling parameters are
\begin{align}
F_{l_1,l_2} &= \int d\boldsymbol{r}\Psi_{l_1}^{*}\left(\boldsymbol{r}\right)h\left(\boldsymbol{r}\right)\Psi_{l_2}\left(\boldsymbol{r}\right),\label{eqpara1}\\
G_{l_1,l_2,l_3,l_4}^{(1)}&= -\int d\boldsymbol{r}_1d\boldsymbol{r}_2\Psi_{l_1}^{*}\left(\boldsymbol{r}_1\right)\Psi_{l_2}^{*}\left(\boldsymbol{r}_2\right)\frac{ke^{2}}{
\left|\boldsymbol{r}_1-\boldsymbol{r}_2\right|}\notag\\
&\quad\cdot\Psi_{l_3}\left(\boldsymbol{r}_1\right)\Psi_{l_4}\left(\boldsymbol{r}_2\right),\label{eqpara2}\\
G_{l_1,l_2,l_3,l_4}^{(2)}&= -\int d\boldsymbol{r}_1d\boldsymbol{r}_2\Psi_{l_1}^{*}\left(\boldsymbol{r}_1\right)\Psi_{l_2}^{*}\left(\boldsymbol{r}_2\right)\frac{ke^{2}}{
\left|\boldsymbol{r}_1-\boldsymbol{r}_2\right|}\notag\\
&\quad\cdot\left[\Psi_{l_3}\left(\boldsymbol{r}_1\right)\Psi_{l_4}\left(\boldsymbol{r}_2\right)-\Psi_{l_4}\left(\boldsymbol{r}_1\right)\Psi_{l_3}\left(\boldsymbol{r}_2\right)\right].\label{hubparameter}
\end{align}
Then the full Hamiltonian can be written
down in the second quantization form as $H'=H_0'+H_C'+H_Z$.

We note that the difference between our configuration interaction method and the
traditional molecular orbital method \cite{Hu.00}
is that here we have transformed the eigenvalue problem with non-orthogonal basis
to a standard eigenvalue problem with orthogonal basis.

\section{Generalized Hubbard model}\label{sec:Hubgen}

We now consider the specific case of a coupled double quantum dot system with each dot capable of holding up to two electrons. Effectively this means that we keep the $s$-orbital only, i.e. $M=2$ and $\nu=1$ in Sec.~\ref{sec:microgen}. The case where the $p$-orbital comes into play will be discussed in Sec.~\ref{subsec:highorbital}. As discussed in our previous publication,\cite{Yang.11} the system can be described by a generalized form of the Hubbard model which retains all terms allowed by symmetry: the total particle number $N$ and the total spin $S_z$ are conserved. The one-body part of the Hamiltonian can be written as:
\begin{equation}
H_0'=\sum_{i\sigma}\left(-\mu_in_{i\sigma}\right)+\sum_{\sigma}\left(-tc_{1\sigma}^\dagger c_{2\sigma}+H.c.\right)
\end{equation}
where $n_{i\sigma}=c_{i\sigma}^\dagger c_{i\sigma}$, $\mu_i$  denotes the chemical potential (or level energy) on site $i$ ($i=1,2$), and $t$ denotes the inter-site hopping (or tunnel coupling).
The Zeeman term is
\begin{equation}
H_Z=\frac{E_{B}}{2}\left(n_{1\uparrow}-n_{1\downarrow}+n_{2\uparrow}-n_{2\downarrow}\right).
\end{equation}

The two-body part $H_C'$ in general includes a Coulomb repulsion term $H_U$:
\begin{equation}
\begin{split}
H_U&=U_1n_{1\uparrow}n_{1\downarrow}+U_2n_{2\uparrow}n_{2\downarrow}+U_{12}(n_{1\uparrow}n_{2\downarrow}+n_{1\downarrow}n_{2\uparrow})\\
&+(U_{12}-J_e)(n_{1\uparrow}n_{2\uparrow}+n_{1\downarrow}n_{2\downarrow}),\end{split}
\end{equation} 
and a term $H_J$ including spin-exchange (denoted by $J_e$), pair-hopping\cite{Slater.36,Anderson.59,Kanamori.63} (denoted by $J_p$), and occupation-modulated hopping terms\cite{Jefferson.96,Hirsch.92} (denoted by $J_t$):
\begin{equation}
\begin{split}
H_J&=-J_ec_{1\downarrow}^\dagger c_{2\uparrow}^\dagger c_{2\downarrow}c_{1\uparrow}-J_pc_{2\uparrow}^\dagger c_{2\downarrow}^\dagger c_{1\uparrow}c_{1\downarrow}\\
&-\sum_{i\sigma}J_{ti}n_{i\sigma}c_{1\overline{\sigma}}^\dagger c_{2\overline{\sigma}}+H.c..\end{split}
\end{equation} 

In the case with non-zero $t$, $U_1$ and $U_2$ but all other parameters ($U_{12}$, $J_e$, $J_p$, $J_{t1}$, $J_{t2}$) being zero, one recovers the form of the usual Hubbard model \cite{Hubbard.63} (on a lattice with only two sites). The inter-site Coulomb repulsion $U_{12}$ was introduced in the study of the charge ordering in strongly correlated materials\cite{Zaanen.96} and the model including the $U_{12}$ term is usually termed as the ``extended Hubbard model''.\cite{Imada.98} In our previous work we have mapped the widely used capacitance model to the $t=0$ case of the extended Hubbard model and have argued that the extended Hubbard model is the minimal model that explains the experiment.\cite{Yang.11}

The spin-exchange and pair-hopping interaction\cite{Slater.36,Anderson.59,Kanamori.63} have been studied in the context of orbital-selective Mott transition.\cite{Liebsch.03}
For atomic orbitals simple relations exist between the parameters (e.g. for $d$-orbitals in free space, $U_1=U_2=U$, $J_e=J_p=J$, $U_{12}=U-2J$ and due to orthogonality $t=J_t=0$). However, in the quantum dot system these relations need not hold since the system usually has much lower symmetry than the free space or a lattice. The occupation-modulated hopping term $J_t$ has not been considered much in the literature except in certain aspects of superconductivity.\cite{Hirsch.92}

The full Hamiltonian is a $16\times16$ matrix for the four-electron two-dot system ($16=4^2$ since each of the two dots allows four possible quantum states). Because the total electron number $N$ and total spin $S_z$ are conserved, it appears in a block-diagonal form. The details are presented in Appendix~\ref{appham}. For a given $(\mu_2,\mu_1)$ one finds the ground state by diagonalizing the Hamiltonian matrix. The charge stability diagram, plotted on a plane with $\mu_2$ and $\mu_1$ as axes, shows how the ground state changes as $\mu_2$ and $\mu_1$ are varied. Experimentally, the chemical potentials of the dots are controlled by the gate voltages $V_R$ and $V_L$ and the charge stability diagram is plotted on the $V_R$-$V_L$ plane. 

In the case of $t=J_e=J_p=J_{t1}=J_{t2}=0$, the ground state can be labeled as $(n_1,n_2)$ since $n_i$ (the electron occupancy on dot $i$) is a  good quantum number. As one or more of the $t$ and $J$ parameters becomes finite, $n_i$ ceases to be conserved and the ground state is a linear combination of  $(n_1,n_2)$ states. The phase boundaries between blocks with different $N$ and $S_z$ are clearly defined as before since there is no mixing between them. Within an ($N$,$S_z$) block we label the charge stability diagram using the $(n_1,n_2)$ state that dominates the ground state. For example, $t$ mixes the (1,0) and (0,1) states. For $\mu_1>\mu_2$ the (1,0) state is the majority state, and vice versa. We then regard the line $\mu_1=\mu_2$ as the phase separator between two regimes in which (1,0) or (0,1) are the majority states.  Experimentally, these boundary lines separating different states within the same $N$ block are diffuse\cite{Simmons.09,Lai.10} due to the hybridization of those states. To fully reproduce this fading effect, a conductance calculation involving excited states is needed.\cite{Cottet.11} In our work we concentrate on the ground state feature so the fading of the boundary lines is beyond the scope of this work.

In this paper we assume $U_1=U_2=U$ (except in the discussion of Fig.~\ref{U1neqU2}), and $J_{t1}=J_{t2}=J_t$. This means that there is a $1,2$-permutation symmetry and the charge stability diagram is symmetric with respect to the line $\mu_1=\mu_2$. Under this assumption plus a condition that $J_t=0$, the charge stability diagram has a particle-hole symmetry (symmetric with respect to the line $\mu_1+\mu_2=U+2U_{12}-J_e$), as can be seen from the matrix form of the Hamiltonian in Appendix~\ref{appham}. Inclusion of $J_t$ effectively changes the values of $t$ in high-occupancy states, thus breaking the particle-hole symmetry. Of course allowing $U_1\neq U_2$ would lead the stability diagram not being symmetric with respect to the line $\mu_1=\mu_2$.

The calculated stability diagram from the generalized Hubbard model will be discussed in the next section. Analysis of the stability diagram shows that  some of the parameters can be extracted directly from experimental plots. Alternatively, all the parameters of the model can be calculated from the microscopic theory [Eqs.~\eqref{eqpara1}, \eqref{eqpara2}, and \eqref{hubparameter}] using a microscopic confinement potential model. Therefore, the generalized Hubbard model establishes a quantitative correspondence between the microscopic theory and the experiments. Since the experiments take $V_R$ and $V_L$ as the basic variable, it is useful to relate $V_R$, $V_L$ to $\mu_2$ and $\mu_1$. In the literature,  $(\mu_2, \mu_1)$ are assumed to be linear combinations of
$\left(V_R,V_L\right)$:\cite{Gaudreau.06,Korkusinski.07}
\begin{align}
\mu_{1} & = \left|e\right|\left(\alpha_{1}V_{L}+\beta_{1}V_{R}\right)+\gamma_1,\label{mu1V}\\
\mu_{2} & = \left|e\right|\left(\alpha_{2}V_{R}+\beta_{2}V_{L}\right)+\gamma_2;\label{mu2V}
\end{align}
where $\gamma_1$ and $\gamma_2$ are constant energy shifts. 

In our previous work\cite{Yang.11} we have presented a mapping between the capacitance model and the extended Hubbard model, which suggests that the coefficients in Eqs.~\eqref{mu1V} and \eqref{mu2V} are related to the parameters by 
\cite{dassarma.Si}
\begin{align}
\alpha_{1} & = \frac{\left(U_{2}-U_{12}\right)U_{1}}{U_{1}U_{2}-U_{12}^{2}},\quad\alpha_{2}=\frac{\left(U_{1}-U_{12}\right)U_{2}}{U_{1}U_{2}-U_{12}^2},
\label{alphabeta}
\end{align}
and $\beta_{1,2}=1-\alpha_{1,2}$.
When the two dots are symmetric, we define $\alpha=\alpha_{1,2}$,  $\beta=\beta_{1,2}$ and Eq.~\eqref{alphabeta} reduces to\cite{Yang.11}
\begin{align}
\alpha=\frac{U}{U+U_{12}},\label{capacalpha}
\end{align}
and $\beta=1-\alpha$.

We also note that the relation implied by Eqs.~\eqref{mu1V} and \eqref{mu2V} is not a unitary transformation. Since we are primarily interested in the lengths of segments on lines $\mu_1-\mu_2={\rm const.}$, the lengths do conserve, up to a factor of electron charge $e$, upon transforming $(\mu_2,\mu_1)\rightarrow(V_R,V_L)$ in our theory. However, one must be cautious in fitting the experiments\cite{dassarma.Si} since the experimental values of the gate voltages should sometimes be rescaled by a factor in order to apply our theory. This uncertainty between the parameter sets $(\mu_2,\mu_1)$ and $(V_R,V_L)$ is unavoidable within the scope of our theory and can only be resolved with precise experimental information.

\section{Results from the generalized Hubbard model}\label{sec:Hubres}

In this section we present charge stability diagrams calculated directly from the generalized Hubbard model. The parameters $U$, $U_{12}$, $t$, $J_e$, $J_p$, and $J_t$ are assumed to be independent of $\mu_2$ and $\mu_1$. This is different from the calculations from the microscopic theory where the overlap integrals of wave functions at different locations on the stability diagram are not guaranteed to be the same and the parameters indeed change slightly at different locations on the charge stability diagram. We also assume $E_B=0$ in this section for simplicity. Throughout this section we discuss the effect of various parameters in the generalized Hubbard model on the charge stability diagram. To facilitate the comparison we plot all figures within the same $\mu_2$, $\mu_1$ range. We also note that although we set the unit of parameters to be meV (so that the results are comparable with experiments and the results from microscopic theory), it can essentially be arbitrarily chosen with the actual energetics being determined by the details of microscopic confinement.

\begin{figure}[]
    \centering
    \includegraphics[width=5.5cm, angle=0]{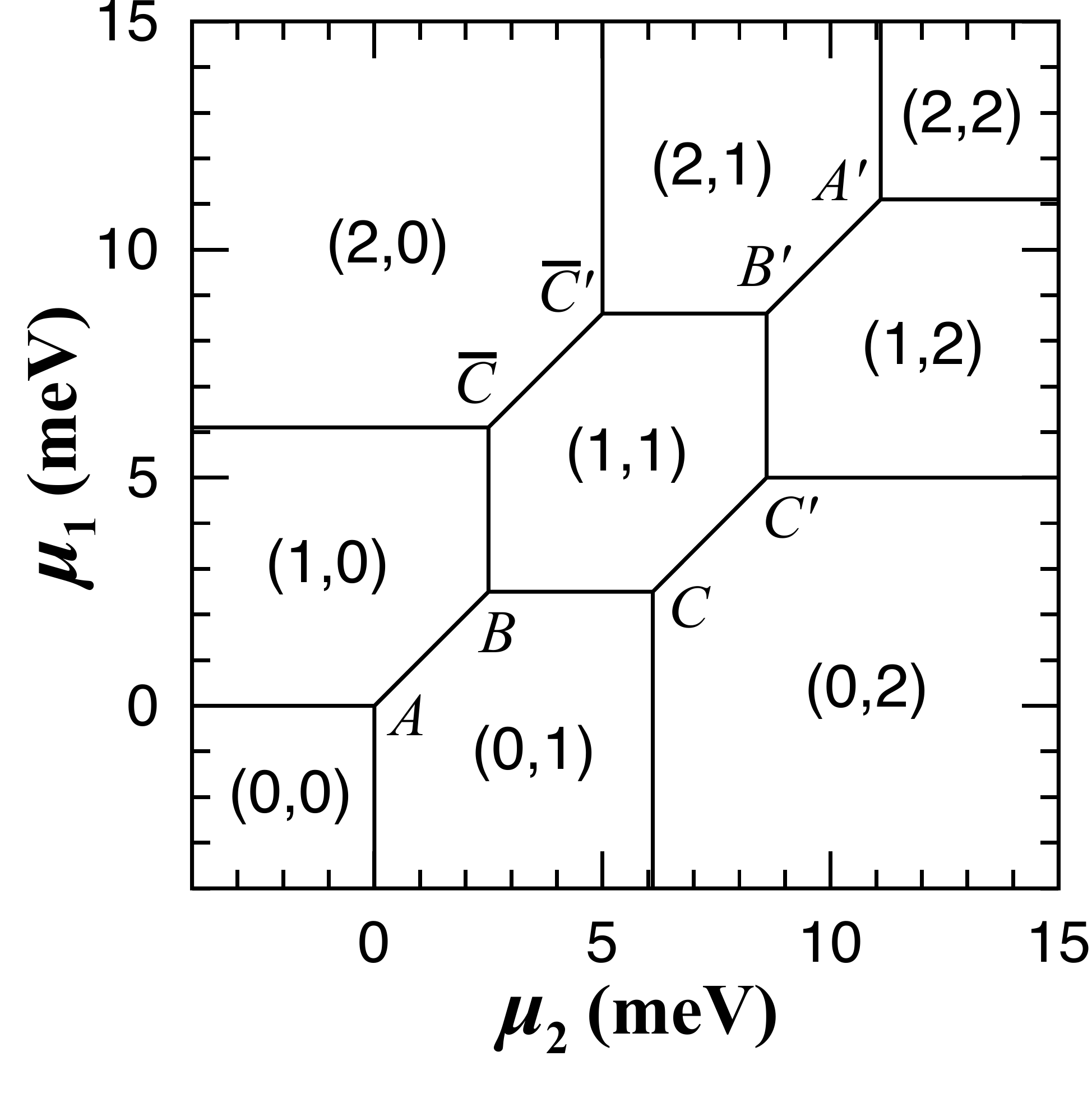}
    \caption{Charge stability diagram calculated at $U=6.1$ meV, $U_{12}=2.5$ meV, and $t=J_e=J_p=J_t=0$.}
    \label{Pettaparam}
\end{figure}

The simplest case is the system of two decoupled quantum dots with on-site Coulomb interaction only, i.e. $U_{12}=t=J_e=J_p=J_t=0$. Its charge stability diagram has a checkerboard pattern which obviously does not fit the experiment. As mentioned in Ref.~\onlinecite{Yang.11}, the extended Hubbard model (with hopping terms neglected) is the minimal model that explains the experiment. Therefore we start with the case $0<U_{12}<U$, while keeping $t=J_e=J_p=J_t=0$. A typical calculation is shown in Fig.~\ref{Pettaparam}. The parameters $U$ and $U_{12}$ are chosen to be the same as Fig.~1 of Ref.~\onlinecite{Yang.11} and the plot is identical to the solid line ($t=0$ result) of Fig.~1(b) in Ref.~\onlinecite{Yang.11}.  

We examine the charge stability diagram in detail, in particular we are interested in the form of the phase boundaries and the coordinates of the ``triple points'',\cite{Wiel.03} defined as the points on the diagram neighboring three different phases. In this case, all eigenenergies are linearly dependent on $\mu_i$, $U$, and $U_{12}$. Therefore all phase boundaries are straight lines, which can be seen clearly from Fig.~\ref{Pettaparam}. The triple points are lettered on the figure with notations that exhibit symmetry considerations. The 1,2-permutation symmetry does two things: First it ensures that triple-points $A$, $B$, $A'$, $B'$ lie on the line $\mu_1=\mu_2$. Second it implies that the coordinates of $\overline{C}$ and $\overline{C'}$ can be found by interchanging indicies 1 and 2 in the corresponding coordinates of $C$ and $C'$. The particle-hole symmetry implies that $\mu_{1,2}(X')=U+2U_{12}-J_e-\mu_{2,1}(X)$, where $X=A, B, C, \overline{C}$. 

The precise coordinates $(\mu_2,\mu_1)$ of the triple points can be readily calculated. We have
$A(0,0)$,
$B(U_{12},U_{12})$,
$C(U,U_{12})$,
and $C'(U+U_{12},2U_{12})$. (Coordinates of other points can be found by symmetry.)
We notice that the length of the line segments $\overline{AB}=\overline{CC'}=\sqrt{2}U_{12}$ and $\overline{BC}=U$. In particular, the length of $\overline{AB}$ and $\overline{CC'}$ does not change upon transforming from $(\mu_2,\mu_1)$ to $(V_R, V_L)$. Therefore from the experimentally measured charge stability diagram,\cite{Petta.05} one can read off the value of $U_{12}$ from the length of the phase boundary between $(1,1)$ and $(0,2)$, as what we have done in Ref.~\onlinecite{Yang.11}. In Ref.~\onlinecite{Yang.11} we have also extracted the value of $U$ from the slope of the phase boundary between $(1,1)$ and $(1,2)$, according to Eq.~\eqref{capacalpha}. We note that the readout of $U_{12}$ is implied by the extended Hubbard model \emph{per se}, while the value of $U$ requires knowledge of the relation between $(\mu_2,\mu_1)$ and $(V_R, V_L)$. While the precise form of this relation is in general unknown, the mapping between the capacitance model and the Hubbard model provides one in the simple cases where all quantum fluctuations vanish. In the cases where the mapping is not necessarily valid, one must exercise caution.

The case $0<U_{12}<U$ considered here is physically reasonable, which can be confirmed by calculations using the microscopic theory (see, e.g. Fig.~\ref{comppotparam}). For completeness we also consider the case with $U_{12}>U$ and the results are shown in Appendix~\ref{extreme}. 

\begin{figure}[]
    \centering
    \includegraphics[width=5.5cm, angle=0]{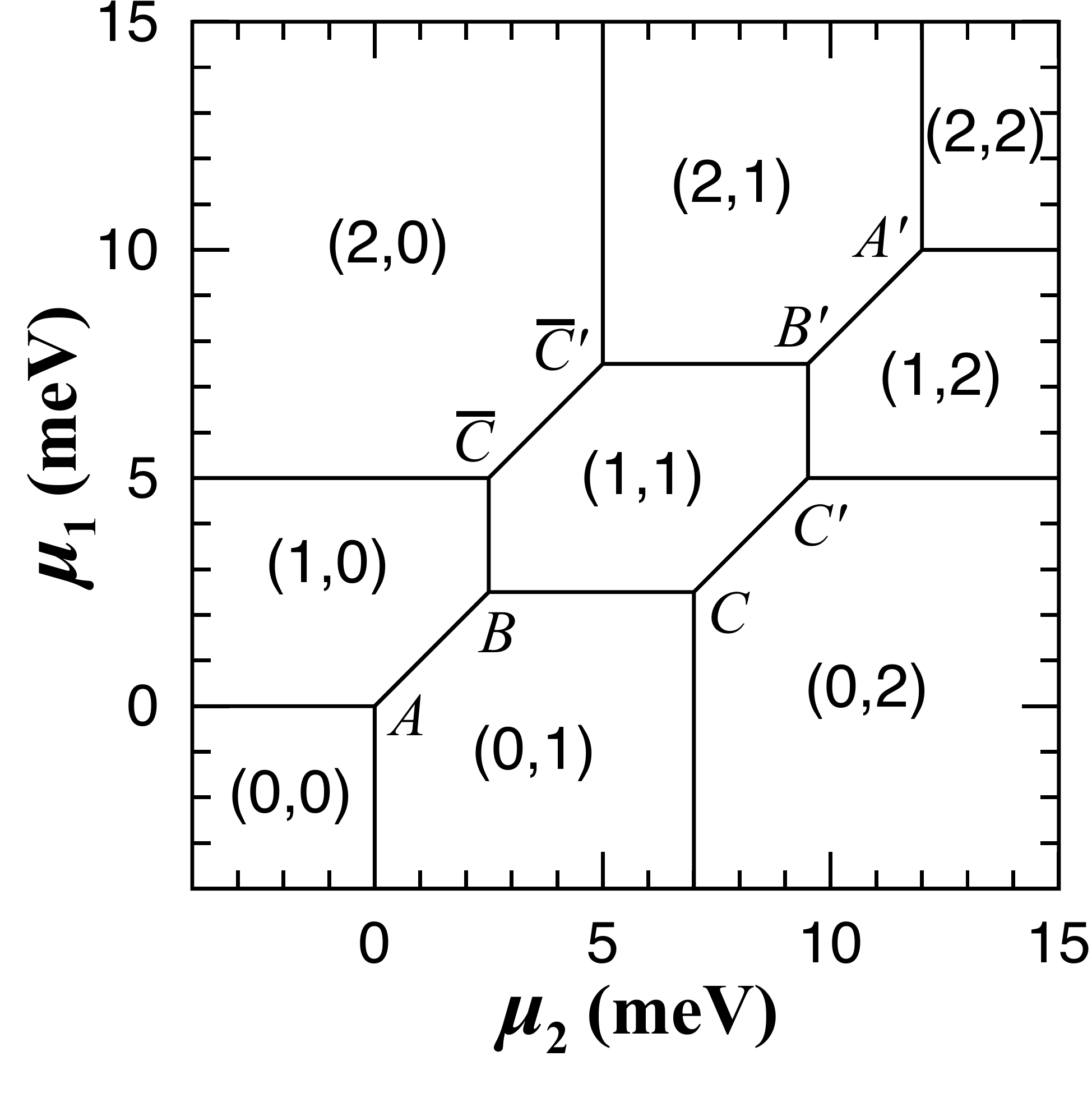}
    \caption{Charge stability diagram calculated at $U_1=5$ meV, $U_2=7$ meV, $U_{12}=2.5$ meV, and $t=J_e=J_p=J_t=0$.}
    \label{U1neqU2}
\end{figure}

In Fig.~\ref{U1neqU2} we show the result calculated allowing $U_1\neq U_2$. Both 1,2-permutation symmetry and particle-hole symmetry are broken. We have the coordinates $(\mu_2,\mu_1)$ of the  triple points as
$A(0,0)$,
$B(U_{12},U_{12})$,
$C(U_2,U_{12})$,
$C'(U_2+U_{12},2U_{12})$,
$\overline{C}(U_{12},U_1)$,
$\overline{C'}(2U_{12},U_1+U_{12})$,
$B'(U_2+U_{12},U_1+U_{12})$, and
$A'(U_2+2U_{12},U_1+2U_{12})$. Note that the lengths of line segments $\overline{AB}$ and $\overline{CC'}$ remain $\sqrt{2}U_{12}$. Since the experimental plot of Ref.~\onlinecite{Petta.05} does not show the whole range of the stability diagram, the explanation of $U_1\neq U_2$ is still consistent with the experiment, i.e., an alternative fit allowing $U_1\neq U_2$ is still valid.

\begin{figure}[]
    \centering
    \includegraphics[width=5.5cm, angle=0]{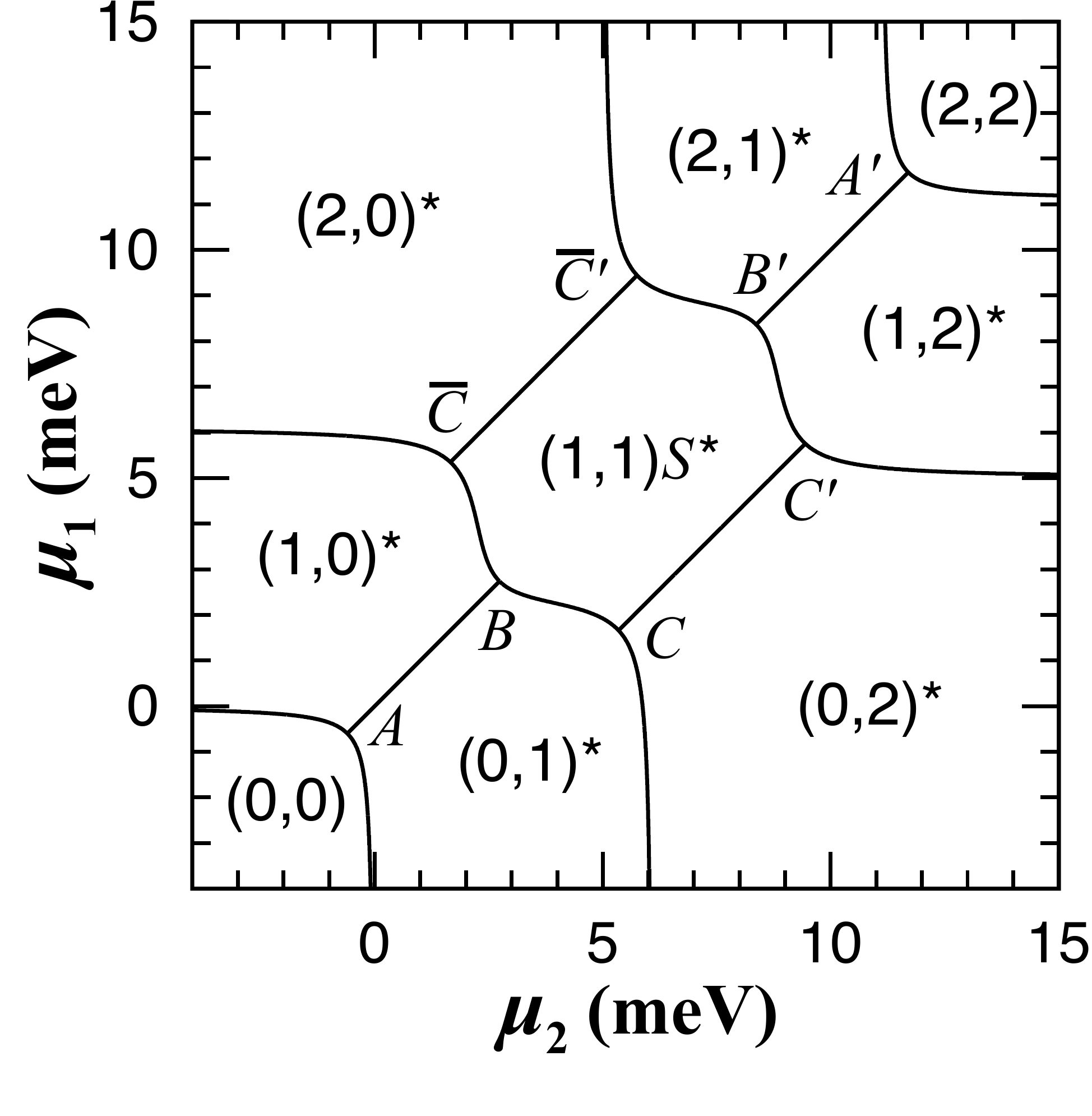}
    \caption{Charge stability diagram calculated at $t=0.6$ meV, $U=6.1$ meV, $U_{12}=2.5$ meV, $J_e=J_p=J_t=0$.}
    \label{t0p6}
\end{figure}

Fig.~\ref{t0p6} presents a typical result calculated with a finite $t$ while $J_e=J_p=J_t=0$. The data shown here is identical to that shown as the red dotted line in Fig.~1(b) of Ref.~\onlinecite{Yang.11}. The main effect of $t$ is to mix states with the same total electron number $N$, except $N=0,4$. In this case one needs to identify the dominant $(n_1,n_2)$ state in the true ground state to label the corresponding region on the charge stability diagram, as mentioned above. In Fig.~\ref{t0p6} we use a star as a superscript of $(n_1,n_2)$ to indicate that this is the majority state in a linear combination rather than the true ground state. Most of the phase boundaries are curved and the one separating $(0,0)$ and $(1,0)/(0,1)$ complex can be identified as hyperbola $\mu_1\mu_2=t^2$. [The boundary between $(2,2)$ and $(2,1)/(1,2)$ complex can be found by particle-hole symmetry condition.] The singlet combination of the $(1,1)$ states dominates the $(1,1)$ component of the stability diagram [denoted by $(1,1)S^*$ in the figure], whose energy eigenvalue can be found by diagonalizing Eq.~\eqref{H5tilde} where simple analytical formula does not exist. The coordinates of $A$ and $B$ can be found analytically: $\mu_2(A)=\mu_1(A)=-t$, $\mu_2(B)=\mu_1(B)=t+(U+U_{12})/2-\sqrt{4t^2+(U-U_{12})^2/4}$, while the coordinates of $C$ and $C'$ cannot be found analytically. It is evident, however, that $\overline{CC'}$ is stretched by the introduction of tunnel coupling $t$.

\begin{figure}[]
    \centering
    (a)\includegraphics[width=5.5cm, angle=0]{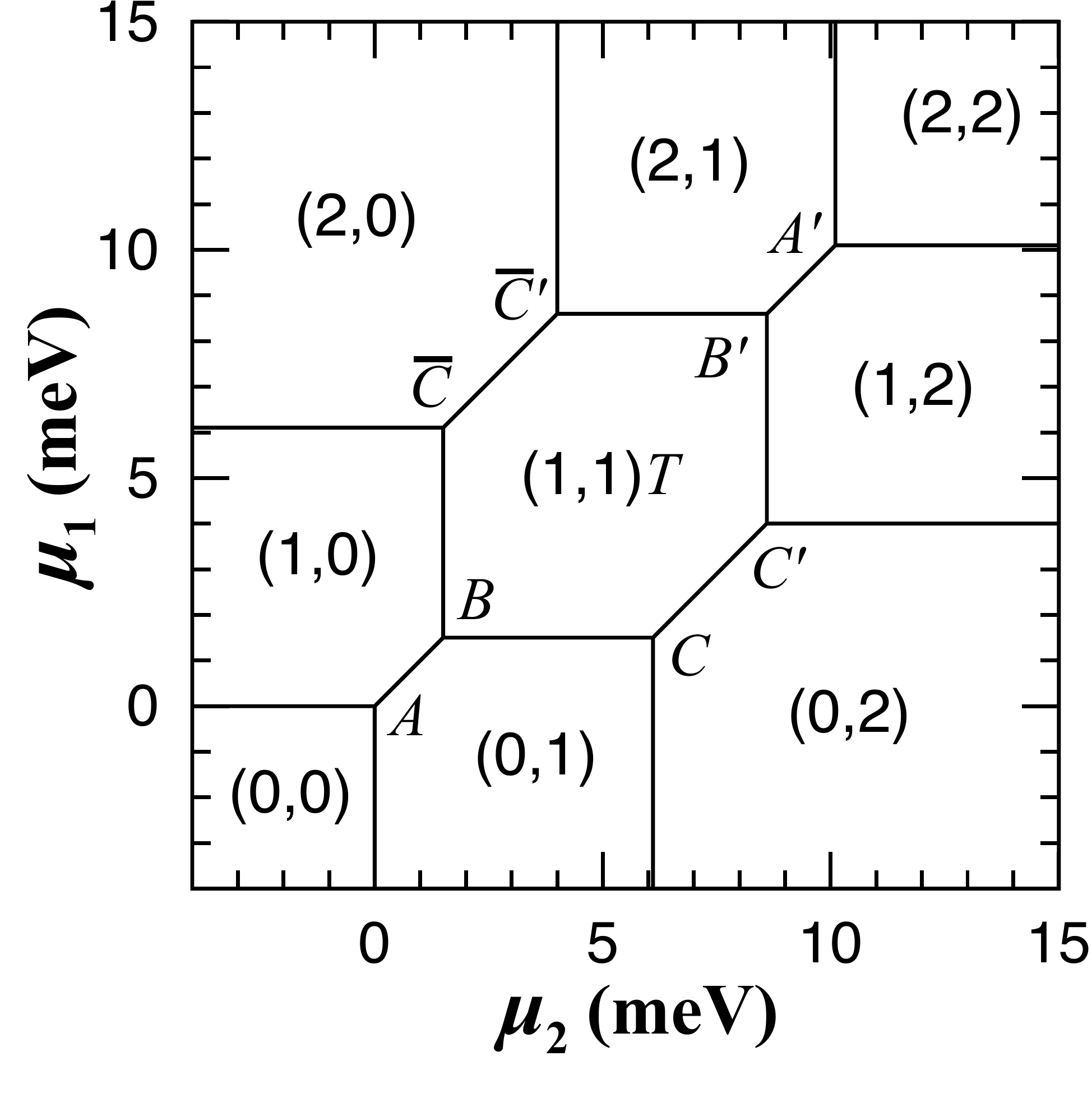}
    (b)\includegraphics[width=5.5cm, angle=0]{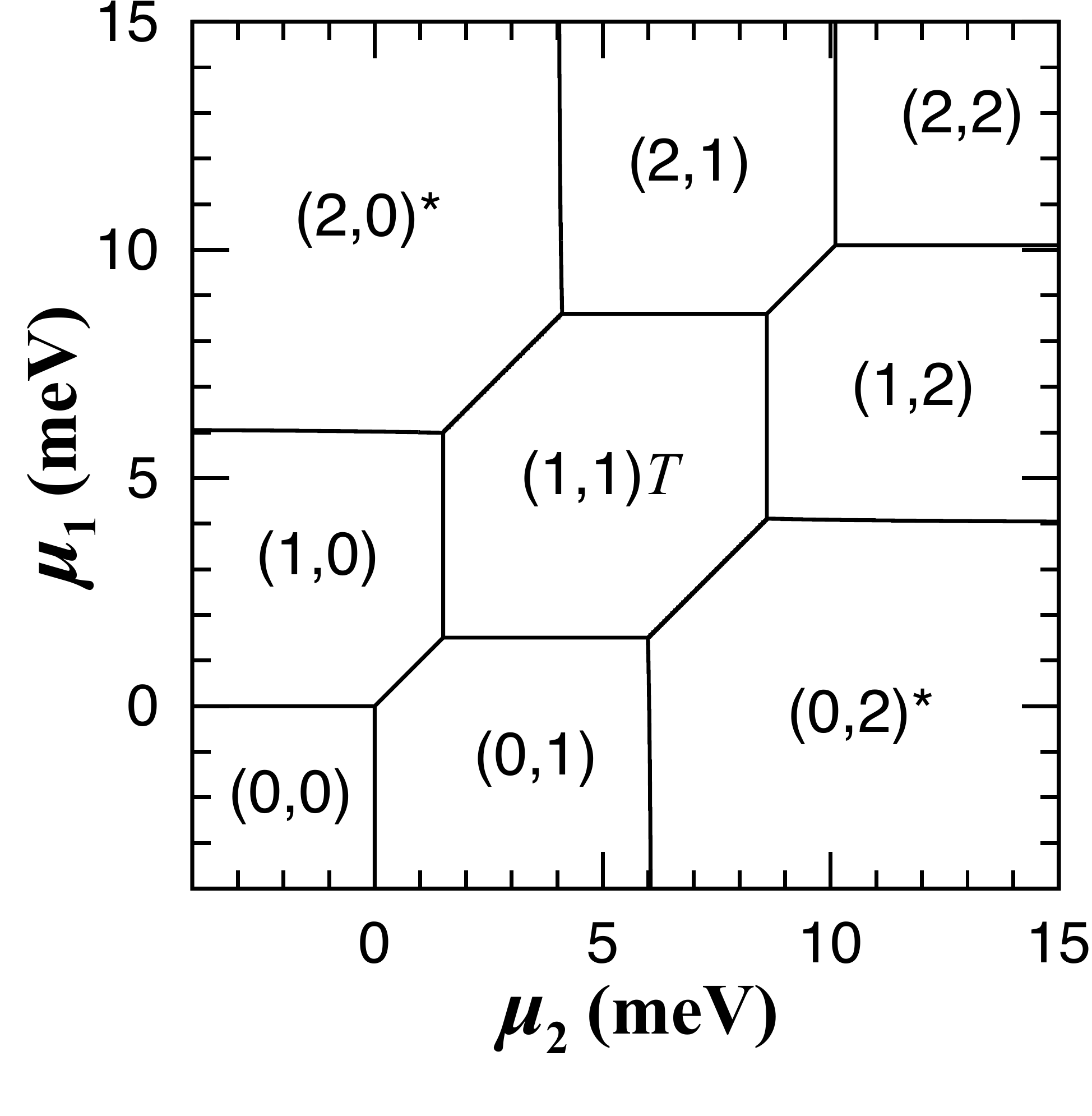}
    \caption{Charge stability diagram calculated at $J_e=1$ meV and (a) $J_p=0$, (b) $J_p=1$ meV. $U=6.1$ meV, $U_{12}=2.5$ meV. $t=J_t=0$.}
    \label{diagJe}
\end{figure}

To understand the effect of $J_e$ we first switch off $t$. Fig.~\ref{diagJe}(a) shows a typical charge stability diagram calculated at $0<J_e<U_{12}$ with $t=J_p=J_t=0$. Since the triplet combination of (1,1) state [labeled by $(1,1)T$] decouples from Eq.~\eqref{H5}, all eigenenergies are again linear functions of $U$, $U_{12}$, and $J_e$ thus the phase boundaries are linear. The coordinates $(\mu_2,\mu_1)$ of the triple points are
$A(0,0)$,
$B(U_{12}-J_e,U_{12}-J_e)$,
$C(U,U_{12}-J_e)$, and
$C'(U+U_{12},2U_{12}-J_e)$.
The length of line segments $\overline{AB}=\sqrt{2}(U_{12}-J_e)$, and $\overline{CC'}=\sqrt{2}U_{12}$. If the whole charge stability diagram is measured, one can read off the value of $J_e$ immediately from the difference between the two segments $\overline{AB}$ and  $\overline{CC'}$ (the effect of other parameters $t,J_t$ must be assumed to be small, though). Typically $J_e$ is smaller than $U$ or $U_{12}$ by at least an order of magnitude. Therefore the $J_e=1$ meV (with $U=6.1$ meV, $U_{12}=2.5$ meV)  shown in Fig.~\ref{diagJe}(a) is slightly exaggerated in order to make its effect clear. In actual systems one expects the effect of $J_e$ on the charge stability diagram to be very small. For a brief discussion of the case $J_e\ge U_{12}$, see Appendix~\ref{extreme}.

The pair hopping interaction mixes the $(2,0)$ and $(0,2)$ states. If $t+J_t\neq0$, it in addition mixes the $(1,1)$ singlet with the (2,0) and (0,2) states, as can be seen from Eq.~\eqref{H5tilde}. Fig.~\ref{diagJe}(b) shows a case with a finite but small $J_p$. Comparison to Fig.~\ref{diagJe}(a) reveals that the change imposed by $J_p$ is very small. The boundaries 
of $(0,2)^*$ and $(2,0)^*$ region are indeed curved but the curvature is hardly detectable. The boundaries between $N=0,1$ blocks and $N=3,4$ blocks remain the same. The correction to the coordinates  of the triple points $C$ and $C'$ (and their mirror $\overline{C}$ and $\overline{C'}$) can be shown to be of order $J_p^2/(U-U_{12})$.

\begin{figure}[]
    \centering
    \includegraphics[width=5.5cm, angle=0]{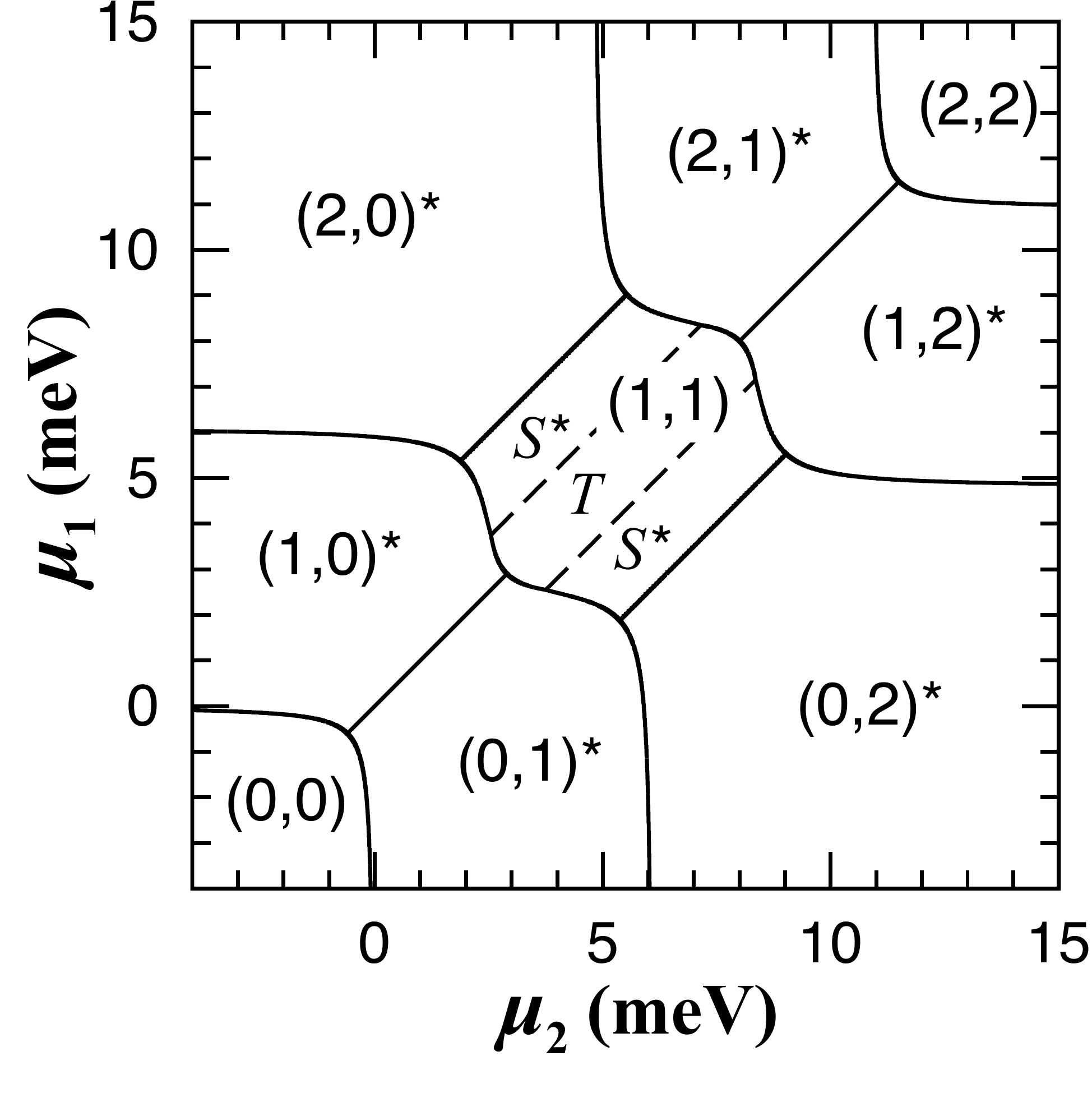}
    \caption{Charge stability diagram calculated at $U=6.1$ meV, $U_{12}=2.5$ meV, $t=0.6$ meV, $J_e=J_p=0.2$ meV. $J_t=0$.}
    \label{STmix}
\end{figure}

Comparison of Fig.~\ref{t0p6} and Fig.~\ref{diagJe}(a) shows that the (1,1) component of the stability diagram has interesting behavior: for $t\gg J_e$ the singlet combination dominates the (1,1) block, while for $t\ll J_e$ the triplet combination dominates. This can be understood from the eigenvalues of Eq.~\eqref{H5}. The triplet has energy $-\mu_{1}-\mu_{2}+U_{12}-J_{e}$ while the singlet has energy $-\mu_{1}-\mu_{2}+U_{12}+J_{e}$ plus contributions due to $t+J_t$. When $t+J_t$ is small the triplet has an energy $2J_e$ lower than the singlet, thus becoming the ground state. However when $t+J_t$ is large the correction due to $t+J_t$ exceeds $2J_e$, making the singlet the dominating ground state. One expects an intermediate value of $(t+J_t)/J_e$ such that the singlet and triplet co-exists. This situation is shown in 
Fig.~\ref{STmix}, where the boundaries separating the singlet and the triplet are plotted as dashed lines.
The condition that the singlet state dominates can be analytically obtained as:
\begin{equation}
(t+J_t)^2>\frac{1}{2}J_e(J_e+J_p+U-U_{12}),
\end{equation}
which is typically satisfied in most microscopic calculations of double quantum dots. The condition that the triplet dominates without having singlet regimes on its sides cannot be expressed analytically.

\begin{figure}[]
    \centering
    (a)\includegraphics[width=5.5cm, angle=0]{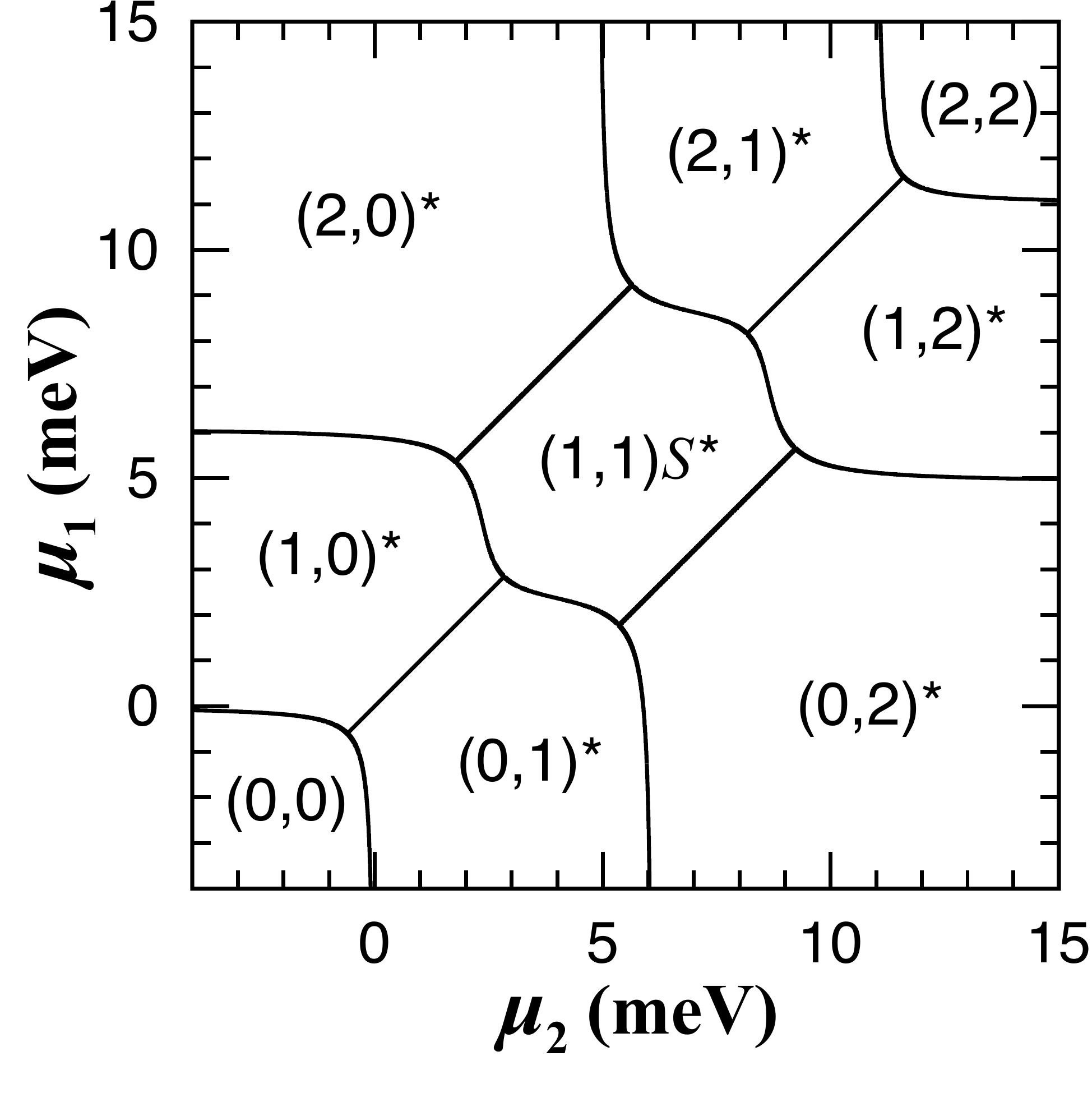}
    (b)\includegraphics[width=5.5cm, angle=0]{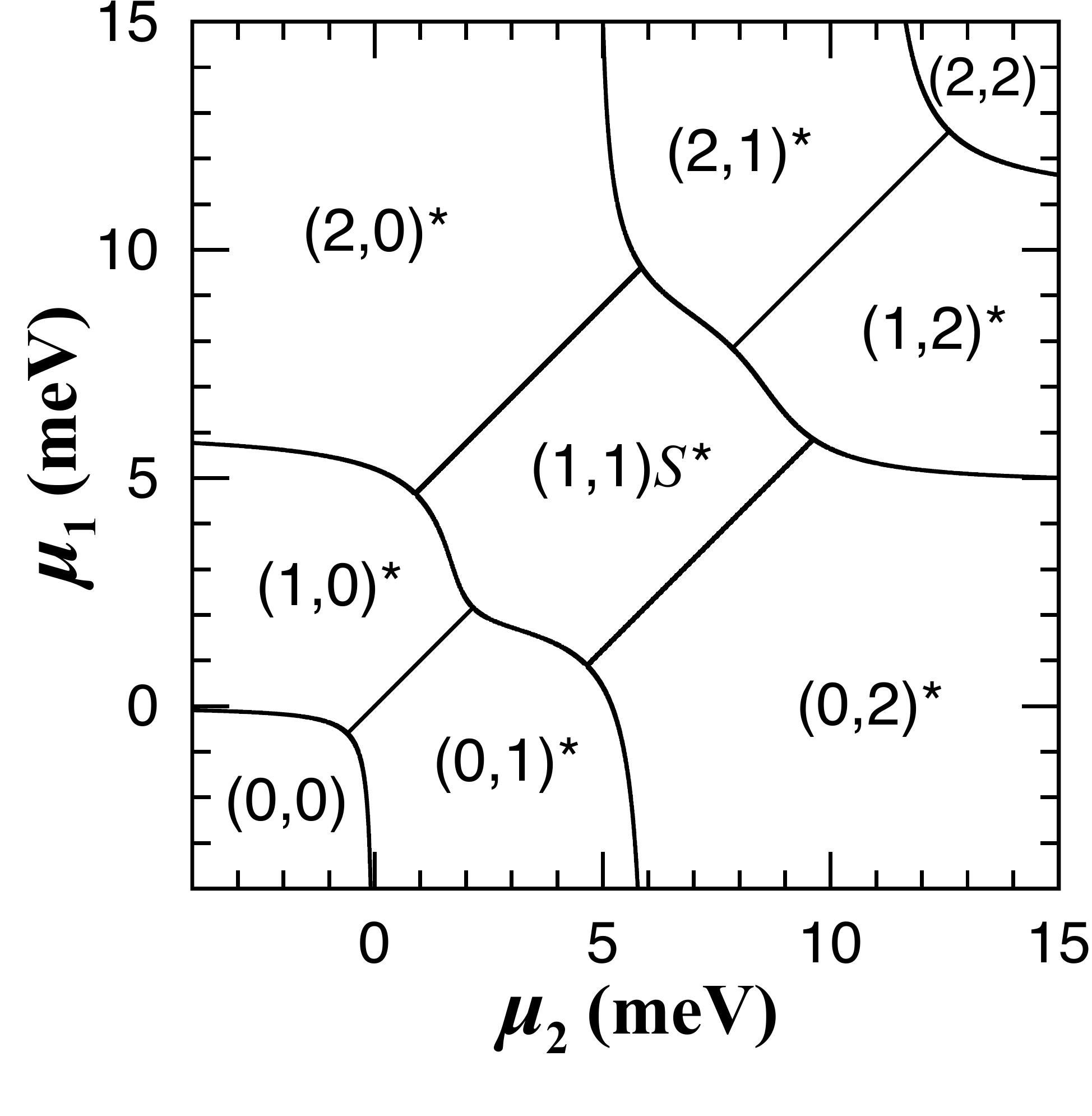}
    \caption{Charge stability diagram calculated at (a) $J_t=0$ (b) $J_t=0.5$ meV  and  $U=6.1$ meV, $U_{12}=2.5$ meV, $t=0.6$ meV, $J_e=J_p=0.1$ meV.}
    \label{allJt}
\end{figure}

At the end of this section we turn on all parameters that we have discussed and consider the effect of $J_t$.
Fig.~\ref{allJt}(a) shows a case with all parameters being non-zero except $J_t$. The figure is essentially very similar to Fig.~\ref{t0p6} since the values of $J_e$ and $J_p$ are very small and have little effect. The particle-hole symmetry is evident. Fig.~\ref{allJt}(b) shows the effect of a relatively large $J_t$. 
First, the $N=0,1$ parts of the stability diagram remain unchanged. Second, the $N=2$ parts [$(0,2)/(1,1)S/(2,0)$ complex] are slightly stretched towards the $N=3$ block. The $N=3$ block [$(2,1)/(1,2)$ complex] is deformed and both of its boundaries are substantially rounded and smoothened. We expect that this is due to the change of an effective $t$  to $t+J_t$ ($N=2$) or $t+2J_t$ ($N=3$). Overall, it is clear that $J_t$ breaks the particle-hole symmetry, making the boundaries between high-occupancy states smoother.

\section{Results from the microscopic theory}\label{sec:microres}

As discussed in Refs.~\onlinecite{Yang.11} and \onlinecite{dassarma.Si}, the microscopic calculation\cite{Burkard.99,Hu.00,Hu.01,Sousa.01,LiQZ.10,Gimenez.07,Nielsen.10,Nielsen.11}
is required to constrain the parameters of the Hubbard model in the physically relevant regime. In other words, although the Hubbard model by itself can have arbitrary parameters, a given physical double-dot system is restricted by the realistic confinement potential which would severely restrict the physical Hubbard parameters for the system.  The details of the application of microscopic theory to our model are described in Sec.~\ref{sec:microgen}. In this section we discuss the application of the microscopic theory to our problem. We shall primarily use two different models of potential: a biquadratic form [Eq.~\eqref{biquadratic}] and a Gaussian form [Eq.~\eqref{gaussian}]. Both of these model potentials are reasonably realistic and used extensively in describing double-dot systems. The biquadratic potential is defined as\cite{LiQZ.10, Helle.05, Pedersen.07}
\begin{equation}
\begin{split}
V_Q(x,y)={\rm Min}\Big\{& \frac{m \omega_0^{2}}{2} [(x+a)^{2}+y^{2}]-\mu_{1},\\
\frac{m \omega_0^{2}}{2} & [(x-a)^{2}+y^{2}]-\mu_{2}\Big\},
\end{split}\label{biquadratic}
\end{equation}
(note that here the two dots are assumed to be symmetric)
and the Gaussian potential reads
\begin{equation}
\begin{split}
V_{G}(x,y)&=-V_1\exp \left[ -\frac{\left( x+a \right)^{2} +y^{2}}{l_{d1}^{2}} \right]\\
-V_2\exp & \left[  -\frac{\left( x-a \right)^{2} +y^{2}}{l_{d2}^{2}} \right] + V_{b} \exp \left[ -\frac{x^{2}+y^{2}}{l_b^2}\right].
\end{split}\label{gaussian}
\end{equation}
Note that Eqs.~\eqref{biquadratic} and \eqref{gaussian} have slightly different forms than that in Ref.~\onlinecite{dassarma.Si}, in order to facilitate the comparison of the effect of different potentials on the charge stability diagram in Sec.~\ref{subsec:diffpot}. This will be explained in Sec.~\ref{subsec:diffpot}.  Besides, we also consider the effects of the additional $p$-orbitals (Sec.~\ref{subsec:highorbital}), and the role of the harmonic oscillator frequency in the calculation with the Gaussian model potential (part of Sec.~\ref{subsec:misc}). All calculations mentioned above are done without magnetic field, and we consider the influence of the magnetic field in Sec.~\ref{subsec:misc}.

The dielectric constant for GaAs $\varepsilon=13.1$ and the effective mass ($m^*=0.067m_e$) are used throughout the microscopic calculation.

\subsection{Influence of model confinement potentials}\label{subsec:diffpot}

\begin{figure}[]
\centering
\includegraphics[width=6.5cm]{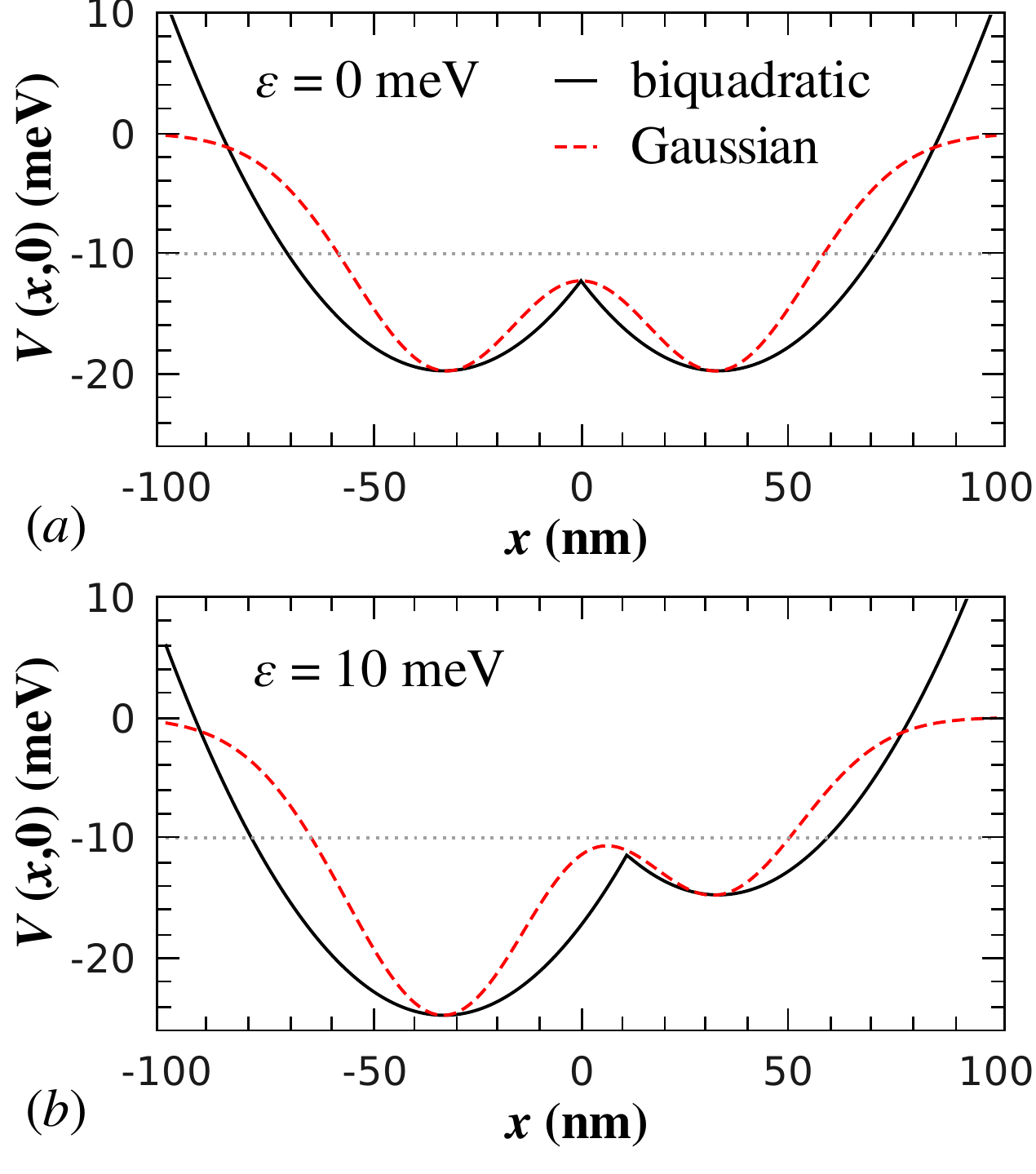}
\caption{(Color online) Profiles of biquadratic (black solid line) and Gaussian (red dashed line) model potentials under two different values of detuning energy $\varepsilon=\mu_2-\mu_1$. For the biquadratic potential, $\hbar \omega_{0}=3.956$ meV, $a=33$ nm. The parameters of the Gaussian potential are derived from the biquadratic one. The gray dotted lines denote the energy reference values of the chemical potential.}
\label{comppot}
\end{figure}

\begin{figure}[]
\centering
\includegraphics[width=8cm]{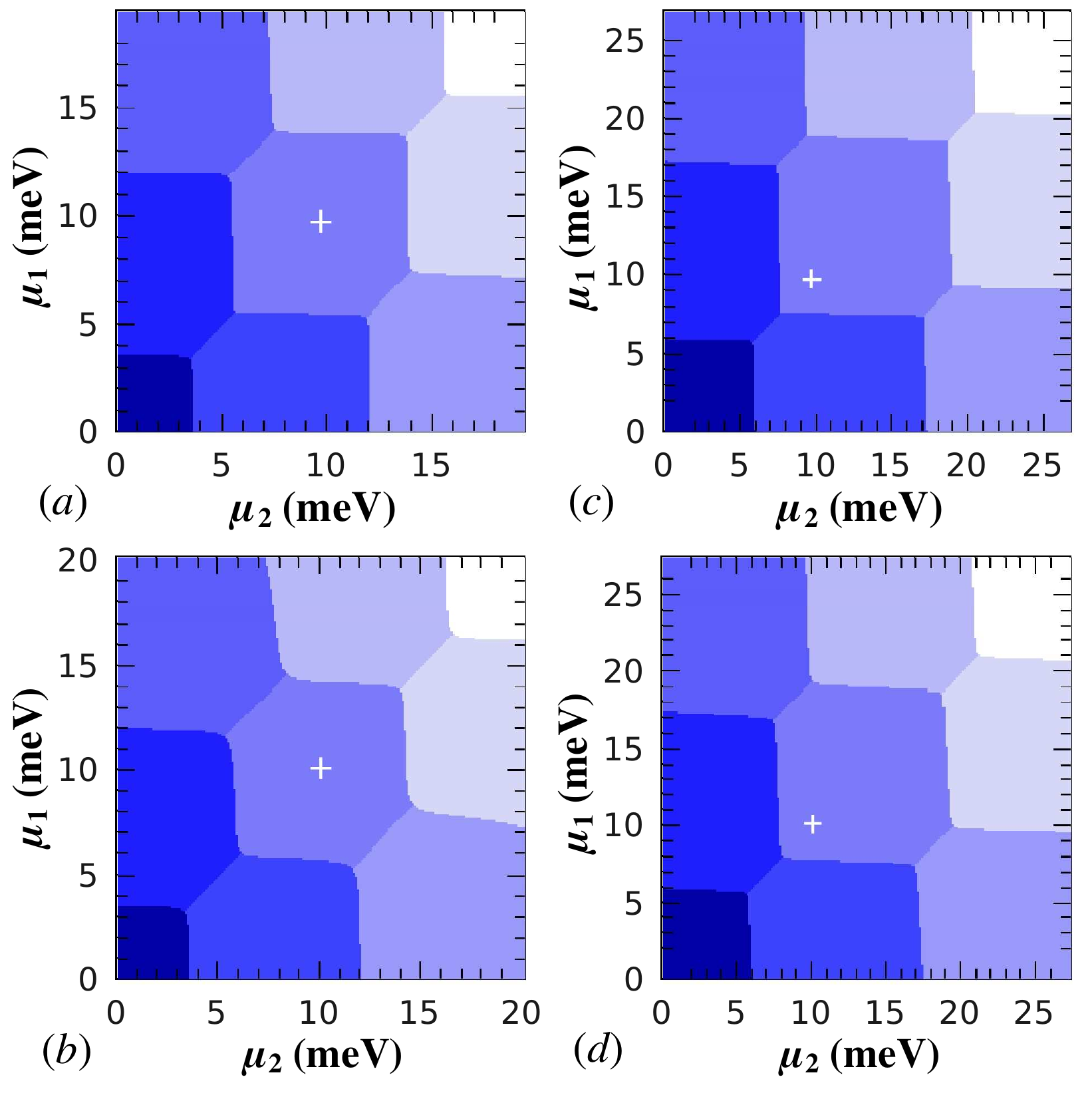}
\caption{(Color online) Charge stability diagrams calculated from biquadratic [panels (a) and (b)] and Gaussian [panels (c) and (d)] model potential different inter-dot distances. The parameters for the biquadratic model are $\hbar \omega_{0}=3.956$ meV, $a=33$ nm (panels (a) and (c)), and $a=28$ nm (panels (b) and (d)). The white crosses [indicating the points at which the comparison of Hubbard parameters (Fig.~\ref{comppotparam}) is conducted] are located at $\mu_{1}=\mu_{2}=9.743$ meV for panels (a) and (c), and $\mu_{1}=\mu_{2}=10.095$ meV for panels (b) and (d).}
\label{SDaQG}
\end{figure}

\begin{figure}[]
\centering
\includegraphics[width=8cm]{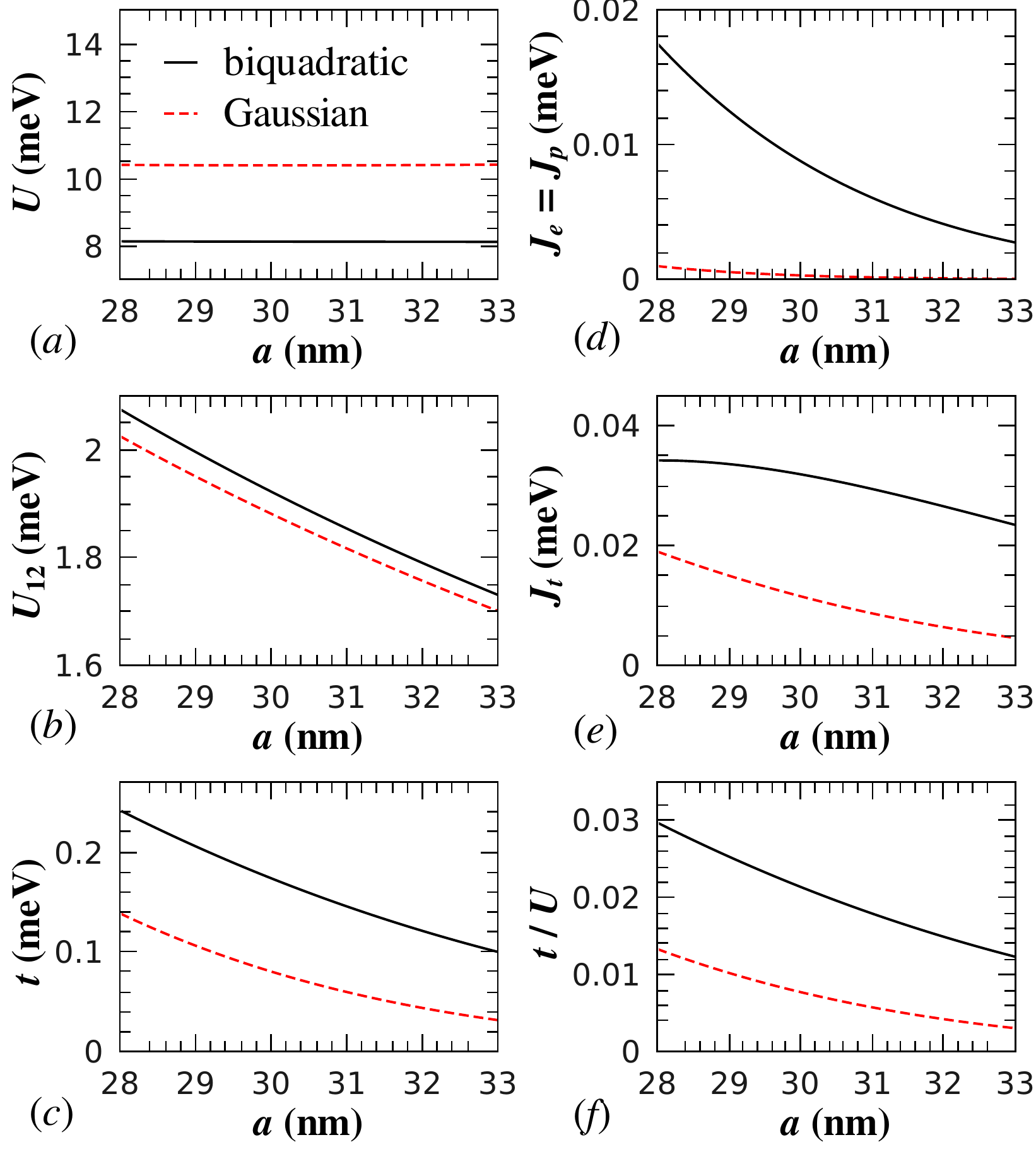}
\caption{(Color online) Parameters in the generalized Hubbard model as functions of half of the inter-dot distance $a$, calculated at the white crosses shown in Fig. \ref{SDaQG}, using both biquadratic and Gaussian potentials.}
\label{comppotparam}
\end{figure}

In the microscopic calculation, the confinement potentials in the quantum dot systems are usually represented by models. Since one has several choices of models, it is useful to understand the consequences of different  model potentials on the charge stability diagram. A similar comparison has been done for the exchange coupling in Ref.~\onlinecite{Pedersen.10}, but in light of the generalized Hubbard model we would like to understand the influence on the overall geometry of the charge stability diagram and the Hubbard parameters. We match the two models of potentials according to their minima and the central barrier (see Fig.~\ref{comppot} for examples at two different values of detuning energy), because these quantities are directly related to the electrostatic potentials that one can experimentally control. The biquadratic model potential is chosen first because it contains only two parameters. The chemical potentials $\mu_{1,2}$ enter the problem in two ways: first, they appear in the formal definition of the model [Eq.~\eqref{biquadratic}]; second, the wave function solution to the Schr\"odinger equation implies the values of the chemical potential [see Eqs.~\eqref{eqhamF1} and  \eqref{eqpara1}]. In our Hund-Mulliken calculation for the biquadratic potential, the chemical potentials defined in these two ways are approximately equal; while for the Gaussian model the chemical potential values calculated from the wave functions are different from $V_1$ and $V_2$ denoted in Eq.~\eqref{gaussian}. Since we want to directly compare the results from the two models under the same electrostatic situations, we tune parameters ($V_{1,2}$) such that the calculated chemical potential values match. This is the main reason that Eq.~\eqref{gaussian} is written in a slightly different way than Ref.~\onlinecite{dassarma.Si}. Since the Gaussian model contains more parameters than the biquadratic model, some of the parameters in the Gaussian model have to be fixed: we fix $l_b=20$nm. The reference energy values (shown as light gray dotted lines in Fig.~\ref{comppot}) correspond to the zero of the chemical potential in Fig.~\ref{SDaQG}. A close examination of Fig.~\ref{comppot} reveals that the main difference between the two models is that the Gaussian model has narrower potential wells. This implies stronger on-site Coulomb interaction and weaker inter-dot tunnelling, as shall be seen in the following discussion.

In Fig.~\ref{SDaQG} we compare the microscopically calculated charge stability diagram for different models of potentials and inter-dot distances. Note that all charge stability diagrams shown in this section have an overall energy shift relative to those shown in Sec.~\ref{sec:Hubres} due to the zero point energy of the harmonic oscillator states. Panels (a) and (b) show the results calculated from the biquadratic potential, while the results from the Gaussian model are shown in panels (c) and (d). The value of $a$ for panels (a) and (c) is $a=33$ nm, while that for panels (b) and (d) is $a=28$ nm. The parameters $a$ and $\omega_0$ (given in the caption) are for biquadratic model, while those for the Gaussian model are derived hereby as explained above. We see that the overall shapes are similar for all panels. Panels (c) and (d) are shown with a larger $x$- and $y$-range than panels (a) and (b), because the Gaussian model has a larger on-site Coulomb interaction than the biquadratic model at the same interdot distance, shifting the phase boundaries to higher energies. This is consistent with the qualitative argument above that the narrower potential wells of the Gaussian model lead to larger on-site Coulomb interactions. Moreover, the rounding effects near the triple points in panels (b) and (d) are more pronounced than panels (a) and (c), which originates from a smaller inter-dot distance $2a$, leading to stronger quantum fluctuations. These arguments will also be confirmed in the calculated Hubbard parameters below. 
We have noted that for the biquadratic model, the chemical potential values derived from Eqs.~\eqref{eqhamF1} and  \eqref{eqpara1} are sometimes slightly different from the values in Eq.~\eqref{biquadratic}. This implies that in Fig.~\ref{SDaQG}(a) and (b) some of the boundary lines, which are parallel to the $x$- or $y$-axis in the Hubbard model calculation, are slightly tilted relative to the axes. This effect is more pronounced for a smaller inter-dot distance 2a [Fig.~\ref{SDaQG}(b)], since a stronger overlap between the wave functions in the two dots leads to a larger deviation of the calculated chemical potential values from the values directly implemented in the model. This tilting effect is not present in the Gaussian model calculation [Fig.~\ref{SDaQG}(c) and (d)], as the chemical potential values are chosen to be the ones calculated from the wave functions.

We would also like to understand the Hubbard parameters as functions of half the inter-dot distance $a$ for the two models.
We choose one particular point on each charge stability diagram in a consistent way to compare the parameters. These points are represented as the white crosses in the (1,1) components of the charge stability diagrams in Fig.~\ref{SDaQG}: they are defined as the mid-way point of the (1,1) segment of line $\mu_1=\mu_2$ for biquadratic model calculations with different values of $a$, and those coordinates are inherited in the Gaussian model calculations, in spite of the fact that they are not in the center of (1,1) regime any more (which is, again, a consequence of the more pronounced localization effect of the Gaussian potential). In Fig.~\ref{SDaQG} the white crosses are actually at the same location for panels (a) and (c), as for panels (b) and (d). We have calculated the Hubbard parameters at these points for a range of $a$ values (which are bracketed by those shown in Fig.~\ref{SDaQG}), and the results are summarized in Fig.~\ref{comppotparam}. Fig.~\ref{comppotparam}(a) reveals that the on-site Coulomb repulsion $U$ is independent of the inter-dot distance, indicating that the wave functions are well localized. $U$ is larger in the Gaussian model than in the biquadratic model by about 35\%, which is a consequence of narrower potential wells in the Gaussian model: the electron wave function is more localized than the biquadratic model. For the same reason, all parameters characterizing the inter-dot interactions are smaller in the Gaussian model than in the biquadratic model, as quantitatively shown in Fig.~\ref{comppotparam}(b)-(f). Fig.~\ref{comppotparam}(b) shows that the inter-dot Coulomb interaction is only slightly decreased in the Gaussian model, while Fig.~\ref{comppotparam}(c) and (e) show that the hopping (tunnel coupling) as well as the occupation-modulated hopping for the Gaussian model is decreased by more than 40\% from that for the biquadratic model. This implies that the difference in the dimensionless ratio $t/U$ can only be more pronounced, as shown in Fig.~\ref{comppotparam}(f). In Fig.~\ref{comppotparam}(d), one sees that the spin-exchange and pair-hopping interaction is substantially smaller in the Gaussian model: it is decreased by an order of magnitude from that for the biquadratic model.

Our results indicate that in modeling the double quantum dot systems, the particular choice of model confinement potential leads to quantitative changes in the Hubbard parameters, although the qualitative behaviors are similar. In a realistic study of these systems, the first-principles calculation of the exact form of the confinement potential from Poisson equation would be required. However, this demands higher precision in both experiments and theoretical calculations, which should be possible in future studies if more precise information about the lithographic gates creating the confinement potential is available.

\subsection{Effect of additional $p$-orbitals}\label{subsec:highorbital}

\begin{figure}[]
\centering
\includegraphics[width=8cm]{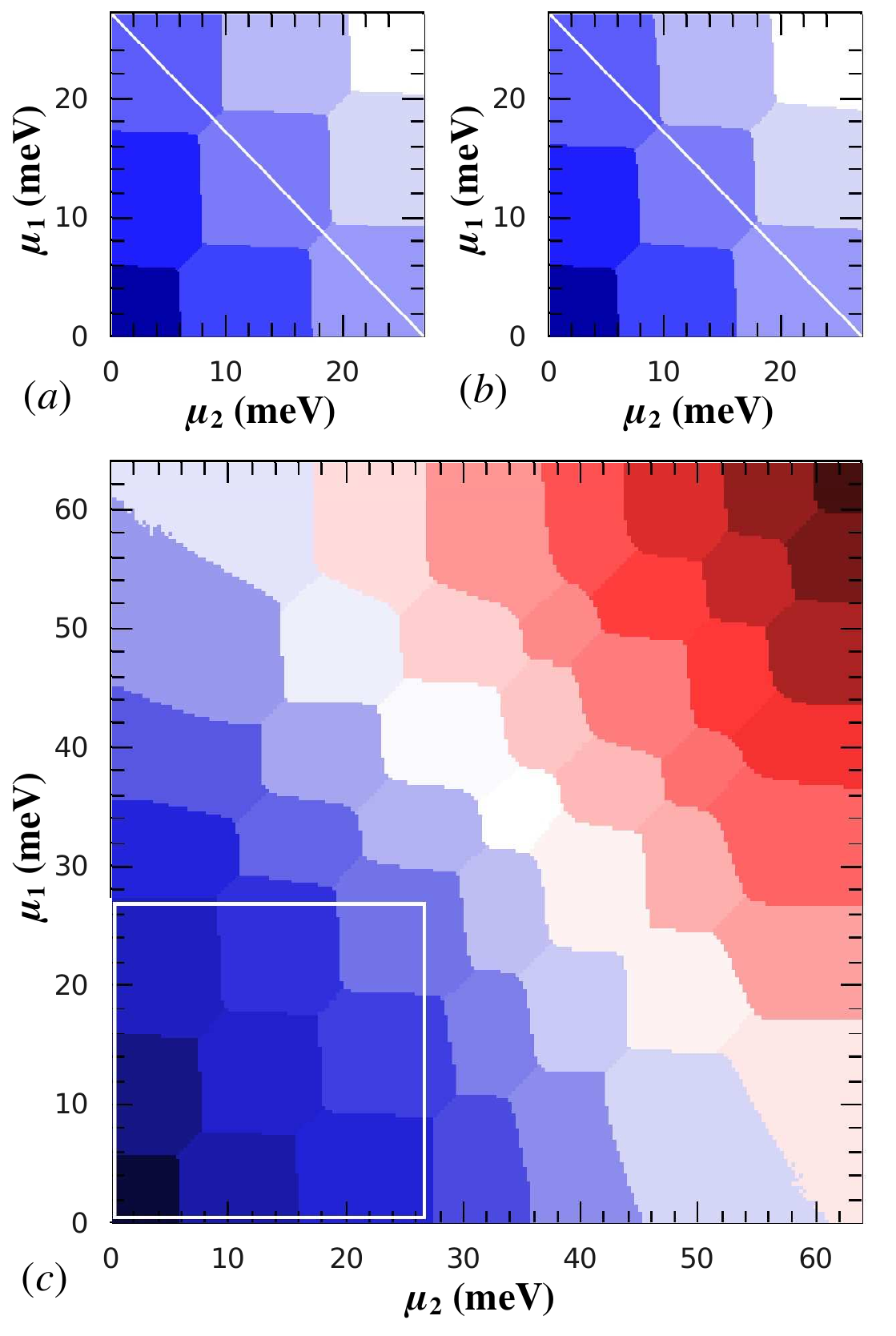}
\caption{(Color online) Charge stability diagrams calculated from the Gaussian potential using (a) $s$-orbitals only and (b) $s$- and $p$-orbitals. The parameters corresponding to the biquadratic potential are $a=30$ nm and $\hbar \omega_{0}=3.956$ meV. Panel (c) shows the full charge stability diagram including all 49 components, while panel (b) shows the stability diagram in the two-electron regime. Panel (b) is a replica of the area enclosed by the white frame in the lower left corner of panel (c).}
\label{spluspdiag}
\end{figure}

\begin{figure}[]
\centering
\includegraphics[width=7cm]{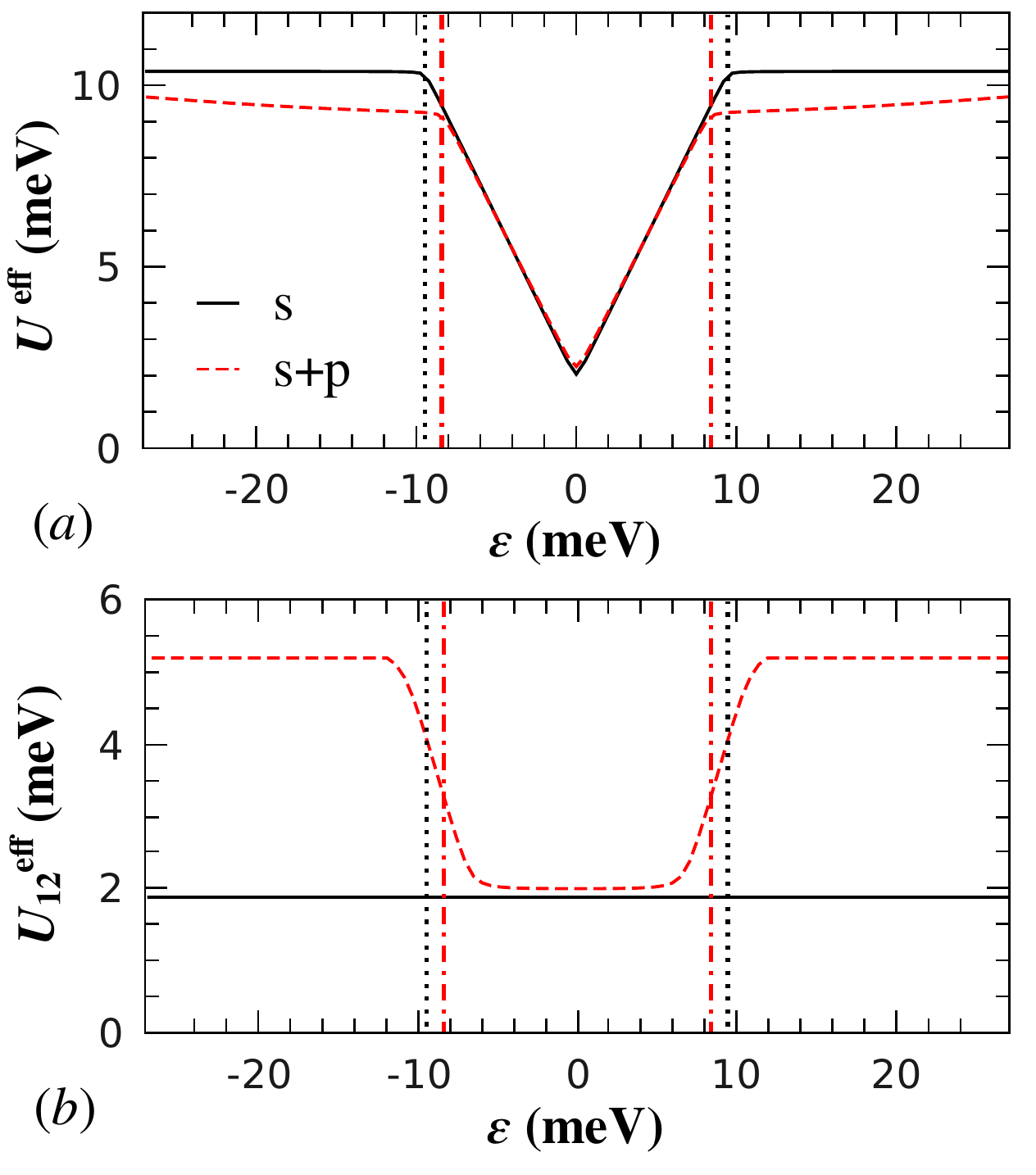}
\caption{(Color online) Effective on-site Coulomb interaction [panel (a)] and inter-site Coulomb interaction [panel (b)] of the results shown in Fig.~\ref{spluspdiag} as a function of the detuning energy $\varepsilon=\mu_2-\mu_1$ calculated keeping $s$-orbitals only (black solid line) and $s$- and $p$- orbitals (red dashed line). The traces where calculations take place are shown as white diagonal lines in Fig.~\ref{spluspdiag}(a) and (b). As $\varepsilon$ is swept from the top left corner to the bottom right corner, the dominating ground state is changed from (2,0) to (1,1), and then to (0,2). The black dotted lines and red dash-dotted lines denote the phase boundaries (where the probabilities of states cross) in the case of $s$-orbitals only and $s$- and $p$-orbitals, respectively.}\label{spluspparam}
\end{figure}

All of the calculations shown in this paper so far, as well as those shown in Refs.~\onlinecite{Yang.11} and \onlinecite{dassarma.Si}, bear the same feature that each quantum dot is only allowed to hold up to two electrons. This means that we have taken $M=2$ and $\nu=1$ in the general discussion in Sec.~\ref{sec:microgen}, keeping the $s$-orbital (denoting the lowest-lying state) only. In the discussion of silicon devices,\cite{dassarma.Si} a multi-electron multi-band situation is encountered. In Ref.~\onlinecite{dassarma.Si} we have reduced the full multi-band problem to an effective two-electron regime and have argued that the multi-band effect can be regarded as a renormalization of Hubbard parameters. In this subsection, we examine this idea in a slightly different context: we consider the additional $p$-orbitals (which are doubly degenerate and can be labeled as $p_x$ and $p_y$ orbitals), the lowest excited states except the $s$-orbitals considered. This can be regarded as taking $M=2$ but $\nu=3$ in Sec.~\ref{sec:microgen}. The full charge stability diagram has 49 components ranging from $(0,0)$ to $(6,6)$. We focus on the part with population up to two electrons per dot $(2,2)$, and study how some of the Hubbard parameters are renormalized by the additional orbitals.

Fig.~\ref{spluspdiag} compares the charge stability diagram calculated keeping $s$-orbitals only and that calculated keeping both $s$- and $p$-orbitals. Panel (a) is the result of keeping the $s$-orbitals only. Panel (c) shows the full charge stability diagram for the case with both $s$- and $p$-orbitals, including all 49 components. Its lower left corner corresponds to the two-electron regime and we replicate the area enclosed by the white frame as panel (b) and compare that to panel (a) side-by-side. Strong similarity between panel (a) and panel (b) is evident and the differences are only in the details.  In Fig.~\ref{spluspdiag}(b), the stability diagram are slightly shifted toward the origin, and the diamond of the $(1,1)$ component shrinks its size. This means that  the Coulomb interaction $U$ is renormalized to a smaller value by the additional orbitals.

To quantify this idea we study the ``effective'' on-site Coulomb interaction ($U^{\rm eff}$) and inter-site Coulomb interaction ($U^{\rm eff}_{12}$) in Fig.~\ref{spluspparam}. The purpose of studying the effective Coulomb interactions is that we want to introduce a method which may be useful for a reduction of the general multi-band problem to the two-electron regime. For the case studied here (retaining three orbitals) it is still possible to calculate all parameters in the generalized Hubbard model, but for a general multi-band problem this calculation is cumbersome and unnecessary, as all the qubit manipulations are done in the two-electron regime and it is sufficient to study the effect of the additional band on this regime.

For a general multi-band problem, the total electron number $N$ and total spin $S_z$ should still be conserved. The full Hamiltonian is block-diagonalized as per $\{N,S_{z} \}$ and we denote the $n^{\rm th}$ lowest energy eigenvalues in each block as $E_{n}^{ \{N,S_{z} \} }$. Then when the $(2,0)$ or $(0,2)$ is the dominant ground state, the effective on-site Coulomb interaction can be defined as the energy difference between the doubly occupied state and the singly occupied state:
\begin{equation}
U^{\mathrm{eff}}=E_{1}^{ \{2,0 \} } - 2 E_{1}^{ \{ 1,\pm \frac{1}{2} \} }.
\label{Uhubeff}\end{equation}
Note that here we have used the fact that the two dots are symmetric. When they are asymmetric, Eq.~\eqref{Uhubeff} gives the Coulomb interaction on the first dot ($U_1$) when calculated at chemical potential values such that (2,0) dominates the ground state, and it gives $U_2$ when calculated in cases where the (0,2) dominates the ground state. 

Moreover, for cases where $(1,1)$ dominates the ground state, the effective inter-site Coulomb interaction $U_{12}^{\mathrm{eff}}$ is defined as the difference between the energy of the state with each dot holding one electron and the total energy of states that one electron is allowed to fill one and the other dot respectively. This means that in the $\{ N=1,S_z=\pm \frac{1}{2} \}$ block both the lowest and the second lowest energy eigenvalues are involved, as we want the electron to occupy a different dot after one of the two dots is occupied. Then  $U_{12}^{\mathrm{eff}}$ is expressed as
\begin{equation}
U_{12}^{\mathrm{eff}}=E_{1}^{ \{2,\pm 1 \} }- E_{1}^{ \{ 1,\pm \frac{1}{2} \} }-E_{2}^{ \{ 1,\pm \frac{1}{2} \} }.
\label{U12hubeff}\end{equation}

In Fig.~\ref{spluspdiag}(a) and (b) we have drawn two diagonal lines in the $N=2$ block of the charge stability diagram, as this is the regime which attracts the most interest. Along these lines we have calculated the effective on-site and inter-site Coulomb interaction as functions of the detuning energy $\varepsilon=\mu_2-\mu_1$, and the results are shown in Fig.~\ref{spluspparam}. The black solid lines in Fig.~\ref{spluspparam} show the case with $s$-orbitals only, corresponding to Fig.~\ref{spluspdiag}(a), while the red dashed lines show the case with $s$- and $p$-orbitals, corresponding to Fig.~\ref{spluspdiag}(b).
One must be cautious when interpreting Fig.~\ref{spluspparam}, as the definitions of the effective Coulomb interaction Eqs.~\eqref{Uhubeff} and \eqref{U12hubeff} is only valid for a selective range of parameters that some particular states dominates the ground state. In Fig.~\ref{spluspparam}(a), the values shown in the range $|\varepsilon|>10$ meV gives the effective on-site Coulomb interaction: For the case with $s$-orbital only, $U^{\rm eff}=10.38$ meV. This is consistent with the value in the Hubbard model $U=10.38$ meV. For the case with both $s$- and $p$-orbitals, $U^{\rm eff}$ changes slightly with the detuning energy but is bound to a narrow range between 9.24 and 9.68 meV. This indicates that the main effect of the $p$-orbital is a renormalization of the effective on-site Coulomb interaction to a slightly smaller value, which has a very weak dependence on chemical potential values. The results shown in the range $|\varepsilon|<10$ meV are meaningless and will not be discussed here. Fig.~\ref{spluspparam}(b) shows the calculated $U^{\rm eff}_{12}$ and the range $|\varepsilon|<6$ meV is directly relevant here. For the case with $s$-orbital only, $U_{12}^{\rm eff}=1.88$ meV, consistent with the value in the Hubbard model $U_{12}=1.88$ meV. For the case with both $s$- and $p$-orbitals, $U_{12}^{\rm eff}=2.00$ meV. Therefore the additional $p$-orbitals does not change effective value of $U_{12}$ appreciably. The calculated values of $U^{\rm eff}$ and $U_{12}^{\rm eff}$ are also consistent with that extracted directly from Fig.~\ref{spluspdiag}(a) and (b). The main result in this subsection is that the effect of additional orbitals can be regarded as a small renormalization of the Hubbard parameters and the main physics is indeed captured in the two-site one-orbital Hubbard model. 

We note, that in Fig.~\ref{spluspparam}(b) the $U_{12}^{\rm}$ value extracted from the range $|\varepsilon|>12$ meV is actually related to the Coulomb interaction between $s$ and $p$ orbitals (which can be denoted as $U_{sp}$), and this information is in principle encoded in the stability diagram of Fig.~\ref{spluspdiag}(c). In fact, the full characterization of the three-orbital charge stability diagram in Fig.~\ref{spluspdiag}(c) shows rich behavior: the components of the stability diagram generically have smaller area for highly occupied states than the one-orbital subspace we are considering, for example the (3,3) and (5,5) states spans a substantially smaller range on the charge stability diagram. This indicates that the on-site Coulomb interaction is weaker for higher-lying orbitals than the $s$ orbitals. In fact, as the $p$ orbitals are populated, the effective central potential barrier between the two dots are effectively lower since $p$ orbitals themselves have higher energies, and the corresponding wave functions are less localized. This naturally leads to the fact that on-site Coulomb interaction for higher-lying orbitals is relatively smaller. Moreover, through the experimental observation of the charge stability diagram, one can in principle infer the effective microscopic confinement potential. The full understanding of this multi-orbital problem 
is, however, beyond the scope of the present work and is an interesting topic for future investigation.

\subsection{Harmonic oscillator frequency in the Gaussian model and the effect of external magnetic field}\label{subsec:misc}

\begin{figure}[tbp]
\centering
\includegraphics[width=8cm]{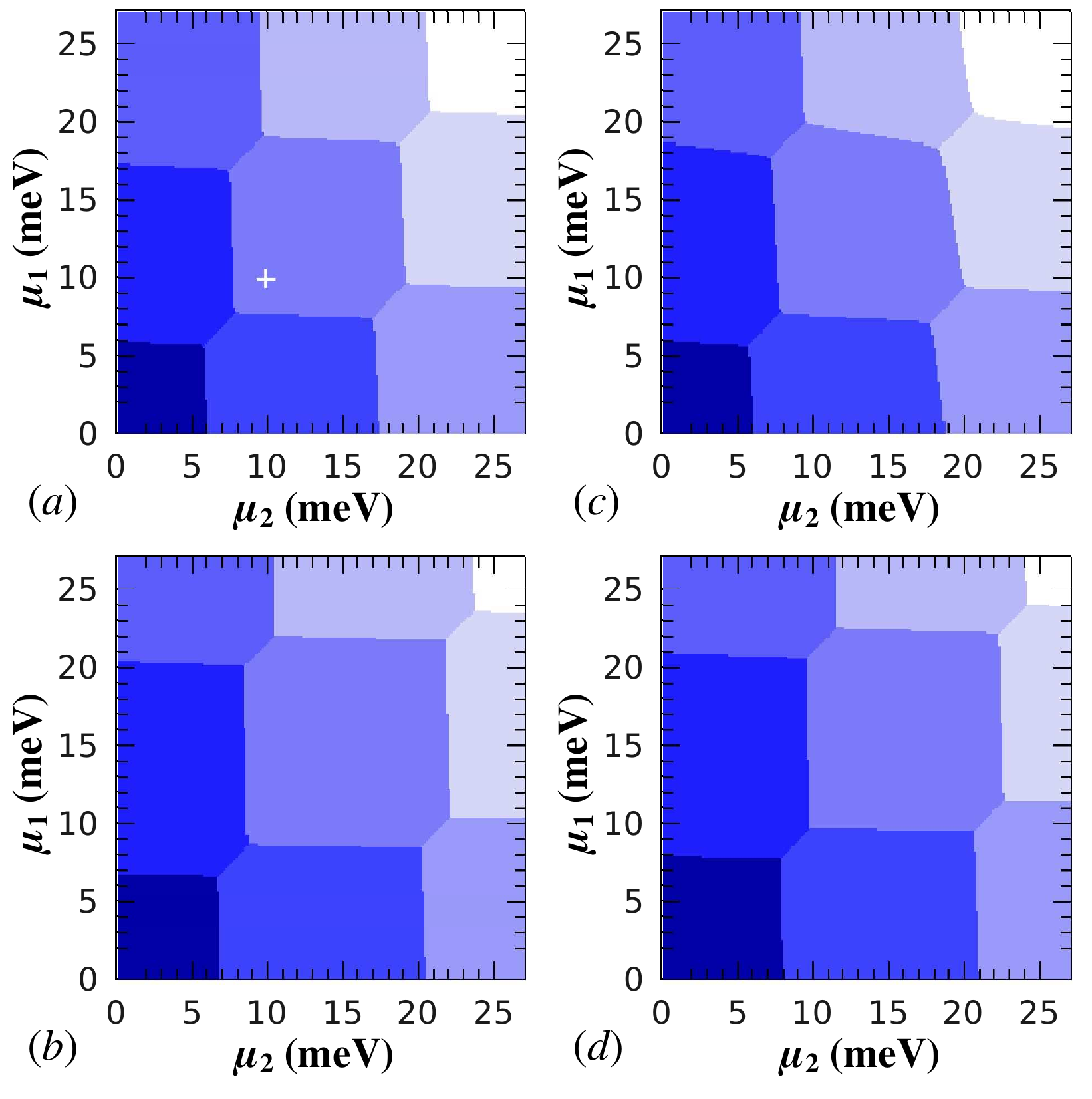}
\caption{(Color online) Charge stability diagrams calculated from Gaussian potential with $a=30$ nm and $\hbar \omega_0=3.956$ meV. (a) $B=0$, the zero-point energies of Fock-Darwin states are chosen according to the microscopic confinement potential at $\mu_1=\mu_2=9.938$ meV (indicated by the white cross) with $\hbar \omega_{0G}=6.487$ meV. (b) $B=0$, the zero-point energies of Fock-Darwin states are $\hbar \omega_{0G}=9.731$ meV. (c) $B=0$, the harmonic oscillator frequencies of the Fock-Darwin states are chosen according to the microscopic potentials for each given $\mu_1$ and $\mu_2$. (d) $B=6$ T, the zero-point energies of Fock-Darwin states are $\hbar \omega_{0G}=6.487$ meV.}\label{figmisc}
\end{figure}

In this subsection we study the role of the harmonic oscillator frequency in the Gaussian model and the effect of external magnetic field.

There are intrinsic difficulties when applying the Gaussian model potential in the Hund-Mulliken approximation: as the chemical potentials change their values, the two potential wells are deformed such that a harmonic-oscillator-type expansion of the confinement minima yields considerably different values of the harmonic oscillator frequencies. Since the harmonic oscillator frequency is the central variable for the Fock-Darwin states, its variance leads to a change in the Hubbard parameters as well as deformation of the charge stability diagram. Moreover, the values of the chemical potentials themselves rely on the calculated Fock-Darwin states, which further depend on how the harmonic oscillator frequency is selected. In our work we have circumvented these difficulties by closely following the biquadratic potential to determine the corresponding parameters of the Gaussian model, as noted in Sec.~\ref{subsec:diffpot}. A fixed harmonic oscillator frequency has been used throughout the calculation for a particular charge stability diagram, which corresponds to a point in the $(1,1)$ regime where the potential wells are symmetric. The motivation of this constraint is that the biquadratic model potential, with $\omega_0$ fixed, has been shown to well describe the experiment with much less number of independent variables than the Gaussian model.\cite{Yang.11,dassarma.Si} We believe that in order to gain physical insight it is sufficient to manipulate the few most important variables. A precise treatment for the Gaussian model, with the harmonic oscillator frequencies appropriately vary with the deformation of the potential well, gives results that are not quite different from the calculations keeping $\omega_0$ fixed in comparison to experiments. However, it is useful to briefly study how the charge stability diagram changes as the harmonic oscillator frequency is varied. 

Fig.~\ref{figmisc}(a) shows the identical result as that shown in Fig.~\ref{spluspdiag}(a). The harmonic oscillator frequency (energy) has been chosen at the locus denoted by a white cross: $\hbar \omega_{0G}=6.487$ meV. In the calculation of Fig.~\ref{figmisc}(b), we artificially multiply the harmonic oscillator energy by a factor of 1.5, thus further localizing the Fock-Darwin wave functions but leaving the confinement potential unchanged. Beside an overall shift in energy (due to the larger $\hbar\omega_0$), the $(1,1)$ component of the charge stability diagram clearly expands its area, meaning that in this case the on-site Coulomb interaction is increased.
In Fig.~\ref{figmisc}(c), the charge stability diagram is calculated using Fock-Darwin states whose harmonic oscillator frequencies are carefully calculated according to the precise form of Gaussian model at each given point on the charge stability diagram. We see that the lower left components ($N<2$) are quite similar to that of Fig.~\ref{figmisc}(a), while for $N>2$ the spatial variation of the on-site Coulomb interaction $U$ is evident. The shape of the (1,1) component of the charge stability diagram is abnormal: it shrinks its area in proximity to the $N=3$ block, while its boundaries with the $N=1$ block remain approximately the same. As a consequence, pairs of the boundaries of the (1,1) regime are not parallel, in contrast to what observed in other calculations and in most experiments. This means that the case considered here, varying the harmonic oscillator frequency of the Fock-Darwin states as the potential wells are changing, is less relevant to experiments that we have considered. The good correspondence between Fig.~\ref{figmisc}(a) and (c) means that the approximation that we have employed is reasonable; Rather, the $\hbar\omega_0$ artificially chosen for Fig.~\ref{figmisc}(b) is too large to accurately reflect the actual solution. 
We therefore conclude, since the present calculations already have quantitative correspondence with experiments, that the method we have used in previous sections using Fock-Darwin states with fixed harmonic oscillator frequency is most suitable for our work.

In Fig.~\ref{figmisc}(d) we show a calculation with parameters same as Fig.~\ref{figmisc}(a) except that there is a finite external magnetic field $B=6$ T. The reason that we group this result with others discussed in this section is that the magnetic field can be regarded to have very similar effect as increasing the harmonic oscillator frequency, as can be seen by similarities between Fig.~\ref{figmisc}(b) and (d). In fact, this can be straightforwardly understood from Eq.~\eqref{eqfrequency}: A finite magnetic field leads to a finite Larmor frequency $\omega_c$, which consequently enters the energy of the Fock-Darwin states in a similar way as the harmonic oscillator frequency (at dot $j$) $\omega_j$. Therefore, a magnetic field effectively localizes the electron wave function, leading to stronger on-site Coulomb interactions and weaker inter-site Coulomb interactions.

\section{Conclusion}\label{conclusion}

\begin{figure}[t]
    \centering
    \includegraphics[width=5.5cm, angle=0]{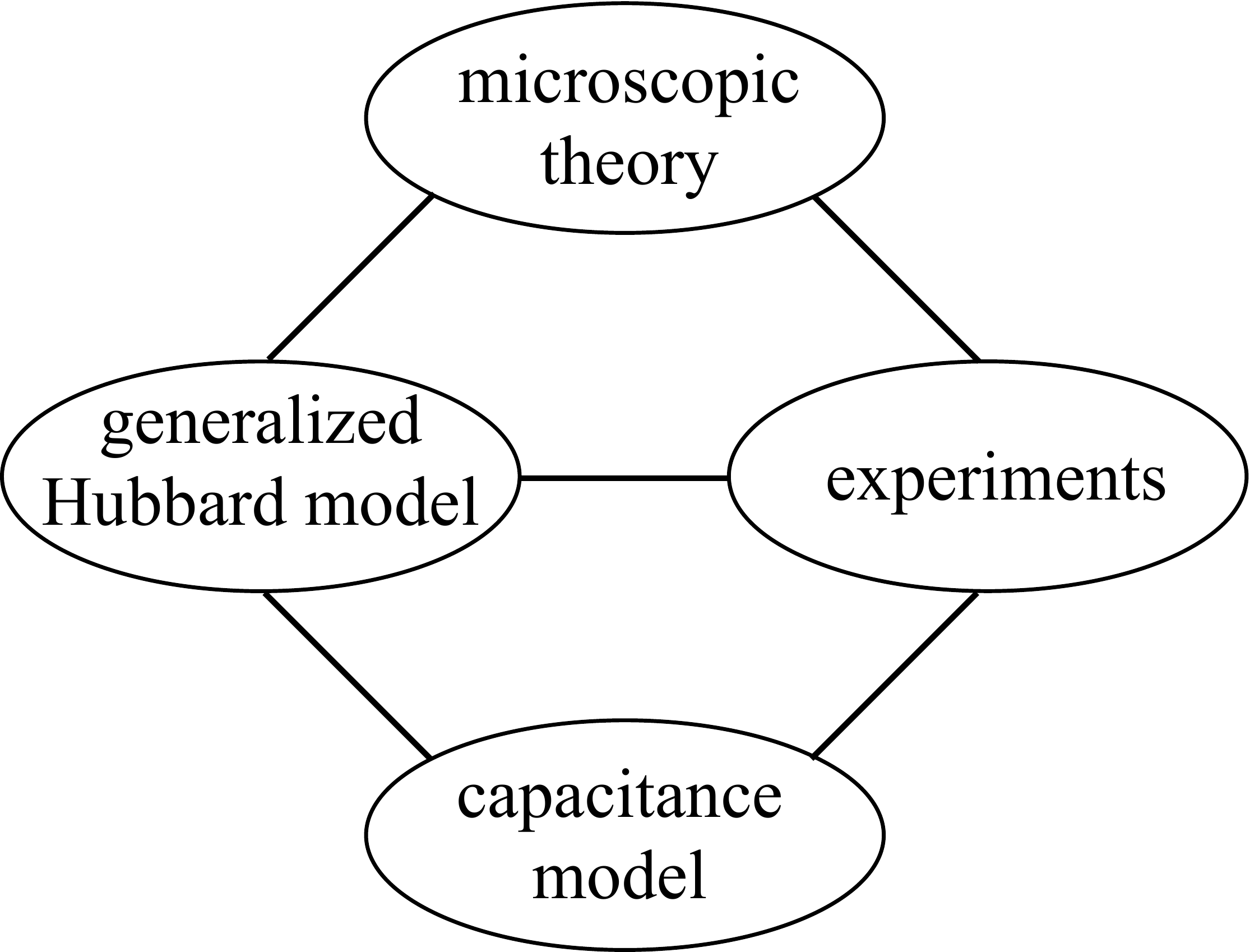}
    \caption{Illustration of the connections between different components discussed in this paper as well as Refs.~\onlinecite{Yang.11} and \onlinecite{dassarma.Si}.}
    \label{triangle}
\end{figure}

In this paper, along with our previous publications Refs.~\onlinecite{Yang.11} and \onlinecite{dassarma.Si}, we have developed a theoretical framework treating the coupled double quantum dot system with a generalized form of Hubbard model. The aim of our work is to establish conceptual and quantitative connections between experiments, existing theoretical approaches and the new theory (see Fig.~\ref{triangle}), thus presenting a unified and transparent description of our current understanding of the double quantum dot system and allowing possibilities for studying more complicated situations in future work. 
The central object of these studies, which is invariably present in all these connections, is the charge stability diagram, the visualization of electron configurations in the double dot system. Given the essential role of semiconductor double quantum systems in solid state quantum computer architectures, our use of the generalized Hubbard model in describing the system is natural since the necessary entanglement effects are embedded in the interaction terms of the Hubbard model.

Experimentally, the charge stability diagram can be directly measured by a quantum point contact,\cite{Elzerman.03} and it is therefore a genuine representation of electron states in the quantum dot system. It contains two kinds of information: the classical Coulomb effect and quantum fluctuations. The classical aspect of the change stability diagram is well understood using the capacitance circuit model.\cite{Beenakker.91,Wiel.03,Hanson.07,Schroer.07} On the other hand, microscopic calculations are usually carried out for other aspects of the quantum dot system\cite{Burkard.99,Hu.00,Hu.01,Sousa.01,LiQZ.10,Gimenez.07,Nielsen.10,Nielsen.11} but only rarely for the charge stability diagram.\cite{ZhangLX.06} This means that our understanding of the quantum aspect of the charge stability diagram has largely been limited to the quantum-mechanical two-level model,\cite{DiCarlo.04,Petta.04,Hatano.05,Huttel.05,Pioro.05} with only a few exceptions.\cite{Gaudreau.06,Korkusinski.07,Cottet.11}

The generalized Hubbard model (see Sec.~\ref{sec:Hubgen}), introduced in Ref.~\onlinecite{Yang.11}, has been naturally written down with only the simplest symmetry considerations in mind. It includes all terms allowed by symmetry, which therefore means that all possible quantum fluctuation effects are included. On one hand, when the quantum fluctuations are completely suppressed, the model is mapped exactly to the capacitance circuit model. On the other hand, quantum fluctuations perturb the charge stability diagram, which is observed both in experiments\cite{Simmons.09,Lai.10} and in theoretical calculations (see Sec.~\ref{sec:Hubres}). The intrinsic assumption of the capacitance circuit model is that the electrostatic energy of a given system are functions of integer electron occupancies on the dots. We have argued that this assumption eliminates the possibility of directly including quantum fluctuations in the capacitance circuit model, as the electron occupancies cease to be good quantum numbers in the presence of quantum fluctuations. Therefore, the generalized Hubbard model is the most direct quantum generalization of the capacitance circuit model.

The generalized Hubbard model contains several parameters. Although their values are not determined by symmetry considerations, they can not be arbitrarily chosen, otherwise unphysical charge stability diagrams may occur (see Appendix~\ref{extreme}). On one hand, we calculate all parameters of the generalized Hubbard model through the microscopic theory, detailed in Sec.~\ref{sec:microgen}. The microscopic calculation involves a numerical solution to the many-electron Schr\"odinger equation under some model confinement potential. The electron wave functions are approximated by linear combinations of the Fock-Darwin states through the configuration interaction method. On the other hand, some of the parameters can be extracted from experimentally measured charge stability diagrams, provided that the resolution is sufficiently high. Therefore the generalized Hubbard model acts as a bridge which quantitatively connects the experiments and the microscopic theory. These considerations, together with the connections to the capacitance model, are shown symbolically in Fig.~\ref{triangle}. Fig.~\ref{triangle} implies that our work in this paper as well as Refs.~\onlinecite{Yang.11} and \onlinecite{dassarma.Si} present a coherent and complete theoretical picture of our current understanding of the double quantum dot system including all effects of quantum fluctuations and entanglement.

In our previous works,\cite{Yang.11,dassarma.Si} we have introduced the generalized Hubbard model, elaborated the motivation and various simplifications, and applied it to explain several specific experiments. We consider the present paper as the concluding one of this three-paper series studying the charge stability diagram of double quantum dot system. In other words, the goal of this paper is to complete the theoretical framework that we have developed, providing detailed information that has not been covered by previous papers,\cite{Yang.11,dassarma.Si} thus forming a comprehensive and coherent picture together with the previous works. First, in Refs.~\onlinecite{Yang.11,dassarma.Si} we have focused on one of the most important quantum parameters in the generalized Hubbard model, namely the tunnel coupling, but other quantum parameters are important as well. In this paper, we have studied the effect of all quantum parameters on the charge stability diagram (Sec.~\ref{sec:Hubres}), and several cases with extreme parameters are shown (Appendix~\ref{extreme}). Second, in Ref.~\onlinecite{Yang.11} we have only used the biquadratic potential as the model confinement potential in the microscopic calculations. Although in Ref.~\onlinecite{dassarma.Si} results from both the biquadratic and Gaussian models are shown, a direct comparison between the two models is missing. In this paper, we have directly compared the results of the charge stability diagram calculated using these two model potentials (Sec.~\ref{subsec:diffpot}). We have found that since the Gaussian model has narrower potential wells, the electron states are more localized, leading to a stronger on-site Coulomb interaction and weaker inter-site quantum fluctuations. Third, in all previous works we have retained the lowest-lying orbital only. In order to justify this simplification, in this paper we have studied the effect of higher-lying orbitals on the one-orbital subspace that we have previously considered (Sec.~\ref{subsec:highorbital}). We have found that the additional $p$-orbitals lead to a small renormalization of Coulomb interaction parameters while leaving the main physics unchanged. This justifies our practice of focusing on the one-orbital subspace. Last, we have studied the role of the harmonic oscillator frequency in the Gaussian model and the effect of external magnetic field (Sec.~\ref{subsec:misc}). We have found that the electron wave functions with fixed value of harmonic oscillator frequency is most suitable for comparison with experiments. The influence of an external magnetic field can be understood as effectively increasing the harmonic oscillator frequency in the Fock-Darwin wave function. Both of these are not covered by our previous works.

As discussed in Sec.~\ref{sec:microgen}, we have neglected the interaction of the quantum dot spin qubit with the environment because environmental decoherence, which is a huge subject by itself, is well beyond the scope of the current paper dealing with the qubit control and charge stability diagram in semiconductor quantum dots. Various environment factors contribute to the decoherence process, which include the charge noise,\cite{Hu.06,Stopa.08} arising from the electromagnetic fluctuations in the surrounding gates and the dynamics in unintentional background impurities,\cite{Gimenez.09,Nguyen.11} the hyperfine coupling to the surrounding nuclear spin bath,\cite{Witzel.06,Cywinski.09,Taylor.07,Reilly.08,Reilly.10} spectral diffusion of the electron spin due to slow nuclear spin dynamics arising from the nuclear spin flip-flops due to dipolar coupling within the nuclear spin bath,\cite{Witzel.05,Witzel.06,Witzel.08} the coupling of electron spins to unintentional background impurity spins in the environment,\cite{Witzel.10} and phonon modes.\cite{Storcz.05,Stavrou.05,Stano.06,Harbusch.10,Hu.11} In many situations, the most important environmental factor is the effect of unintentionally introduced background charged impurities, since depending on the proximity between the impurities and the qubit, the interaction with the impurities could be of the same order of magnitude as the on-dot electronic interactions as both are mediated by the long-range Coulomb coupling.\cite{Gimenez.09,Nguyen.11}  The environmental nuclear spin bath affects the electron spin both through a ``direct'' hyperfine interaction between the spin qubit and the nuclei, and an ``indirect'' coupling due to the spectral diffusion process resulting from the dipolar interaction between the nuclei. Recent experimental progresses\cite{Bluhm.11,Foletti.09} have demonstrated control over the nuclear spin bath, making the nuclear spin bath less destructive and the effect of the impurities more prominent. In some systems such as Si or C, the direct hyperfine coupling can be eliminated by using isotropic purification of the nuclear elements.\cite{Witzel.10} Phonon modes are the least important environmental factor for spin qubits working at low ($\sim100$ mK) temperatures: Firstly, the coupling between the spin qubit and the phonon modes are orders of magnitude weaker than the effect of charged impurities and nuclear spin bath mentioned above. Secondly, the effect of phonon modes can be suppressed by cooling the system down to very low temperatures, while the effects due to charged impurities and the nuclear spin bath remain important even at low temperatures. In this paper, we focus on the electronic interaction within the quantum dot system and study its consequence on the charge stability diagram, and quantum decoherence due to environmental factors is beyond the scope of the current work.

Our work suggests several possible directions for future studies. First, it is useful to go beyond the model potential and Hund-Mulliken approximation, and develop a first-principles Poission-Schr\"odinger approach for a realistic calculation of quantum dot confinement. This involves both extracting the detailed form of confinement potentials from experimentally fabricated devices, and an exact treatment of Schr\"odinger equation under this realistic confinement potential. However, this would require a precise experimental control of disorder and impurities during the fabrication, as well as the knowledge of the underlying Fermi sea. Moreover, the calculation expenses are much higher than what we have done in this work, although they are not necessarily prohibitive since modern numerical techniques, e.g. density functional theory, have already demonstrated their power in successfully explaining many aspects of complex materials as well as nanoscale systems. Second, although we have argued that the additional higher-lying orbitals pose no more than a renormalization of Coulomb parameters in the single-orbital subspace, the issue for the full multi-orbital multi-electron problem may not yet be settled. A detailed study of the realistic multi-orbital multi-electron problem would in particular greatly benefit our understanding of the silicon quantum dot system\cite{dassarma.Si} as this multi-electron effect may be one of the obstacles of realizing and manipulating the singlet/triplet qubit in silicon double quantum dot system.
Last, we have focused on equilibrium properties of an isolated quantum dot system where the total electron number and the total spin are conserved. However, non-equilibrium effects are intertwined with the measurement of the charge stability diagram as well as the manipulation of qubits. The charge stability diagram is measured through a tunnelling current, which itself is a non-equilibrium property. In Ref.~\onlinecite{Cottet.11} a study of the charge stability diagram in terms of the microscopic conductance(admittance) is given, and much of our theory can also be applied along this direction. Of course, such a dynamical theory would be much more complicated than what we have studied here, since the excited states must be considered in all non-equilibrium calculations. Moreover, the coupling of the quantum dot system to its environment, which has been neglected in this paper, is by no means insignificant. 
Therefore, the assumptions that we have made in this paper as well as Refs.~\onlinecite{Yang.11,dassarma.Si}, that the system is isolated and in equilibrium, should be lifted in future studies by extending the generalized Hubbard model.
The theoretical approach that we have presented here provides a foundation for these future research directions.

\section*{Acknowledgements}
We thank J. P. Kestner for helpful discussions.
This work is supported by IARPA and LPS.

\appendix
\renewcommand{\theequation}{A-\arabic{equation}}
\setcounter{equation}{0}
\section{Matrix form of the generalized Hubbard model}\label{appham}

The full Hamiltonian is a $16\times16$ matrix. Because the total particle number ($N$) and total spin ($S_z$) are conserved, it appears in a block-diagonal form, as follows.

(1) $N=0$, $S_z=0$. On the basis $\left|0\right\rangle$,
\begin{equation}
H_{1}=0.
\end{equation}

(2) $N=1$, $S_z=\frac{1}{2}$. On the basis $\left\{ c_{1\uparrow}^{\dagger}\left|0\right\rangle, c_{2\uparrow}^{\dagger}\left|0\right\rangle \right\} $,
\begin{equation}
H_{2}=\left(\begin{array}{cc}
-\mu_{1}+E_{B} & -t\\
-t & -\mu_{2}+E_{B}\end{array}\right).
\end{equation}

(3) $N=1$, $S_z=-\frac{1}{2}$. On the basis $\left\{ c_{1\downarrow}^{\dagger}\left|0\right\rangle, c_{2\downarrow}^{\dagger}\left|0\right\rangle \right\} $,
\begin{equation}
H_{3}=\left(\begin{array}{cc}
-\mu_{1}-E_{B} & -t\\
-t & -\mu_{2}-E_{B}\end{array}\right).
\end{equation}

(4) $N=2$, $S_z=1$. On the basis $c_{1\uparrow}^{\dagger}c_{2\uparrow}^{\dagger}\left|0\right\rangle$ (labeled as $T_+$),
\begin{equation}
H_{4}=-\mu_{1}-\mu_{2}+U_{12}-J_{e}+2E_{B}.
\end{equation}

(5) $N=2$, $S_z=0$. On the basis $\left\{ c_{1\uparrow}^{\dagger}c_{2\downarrow}^{\dagger}\left|0\right\rangle, c_{2\uparrow}^{\dagger}c_{1\downarrow}^{\dagger}\left|0\right\rangle, c_{1\uparrow}^{\dagger}c_{1\downarrow}^{\dagger}\left|0\right\rangle, c_{2\uparrow}^{\dagger}c_{2\downarrow}^{\dagger}\left|0\right\rangle \right\} $,
\begin{widetext}
\begin{equation}
H_{5}=\left(\begin{array}{cccc}
-\mu_{1}-\mu_{2}+U_{12} & J_{e} & -(t+J_{t1}) & -(t+J_{t2})\\
J_{e} & -\mu_{1}-\mu_{2}+U_{12} & -(t+J_{t1}) & -(t+J_{t2})\\
-(t+J_{t1}) & -(t+J_{t1}) & -2\mu_{1}+U_1 & J_{p}\\
-(t+J_{t2}) & -(t+J_{t2}) & J_{p} & -2\mu_{2}+U_2\end{array}\right).\label{H5}
\end{equation}
Note that the linear combinations $\left(c_{1\uparrow}^{\dagger}c_{2\downarrow}^{\dagger}-c_{2\uparrow}^{\dagger}c_{1\downarrow}^{\dagger}\right)/\sqrt{2}$ (labeled as $T_0$) and $\left(c_{1\uparrow}^{\dagger}c_{2\downarrow}^{\dagger}+c_{2\uparrow}^{\dagger}c_{1\downarrow}^{\dagger}\right)/\sqrt{2}$ (labeled as $S$) decouple the $T_0$ state with energy $-\mu_{1}-\mu_{2}+U_{12}-J_{e}$, thus forming a triplet which degenerate at zero magnetic field. The remaining $3\times3$ matrix reads
\begin{equation}
\widetilde{H}_{5}=\left(\begin{array}{ccc}
-\mu_{1}-\mu_{2}+U_{12}+J_{e} & -\sqrt{2}(t+J_{t1}) & -\sqrt{2}(t+J_{t2})\\
-\sqrt{2}(t+J_{t1}) & -2\mu_{1}+U_1 & J_{p}\\
-\sqrt{2}(t+J_{t2}) & J_{p} & -2\mu_{2}+U_2\end{array}\right).\label{H5tilde}
\end{equation}

(6) $N=2$, $S_z=-1$. On the basis $c_{1\downarrow}^{\dagger}c_{2\downarrow}^{\dagger}\left|0\right\rangle$ (labeled as $T_-$),
\begin{equation}
H_{6}=-\mu_{1}-\mu_{2}+U_{12}-J_{e}-2E_{B}.
\end{equation}

(7) $N=3$, $S_z=\frac{1}{2}$. On the basis $\left\{ c_{1\uparrow}^{\dagger}c_{2\uparrow}^{\dagger}c_{1\downarrow}^{\dagger}\left|0\right\rangle, c_{1\uparrow}^{\dagger}c_{2\uparrow}^{\dagger}c_{2\downarrow}^{\dagger}\left|0\right\rangle \right\} $,
\begin{equation}
H_{7}=\left(\begin{array}{cc}
-2\mu_{1}-\mu_{2}+U_1+2U_{12}-J_{e}+E_{B} & -(t+J_{t1}+J_{t2})\\
-(t+J_{t1}+J_{t2}) & -2\mu_{2}-\mu_{1}+U_2+2U_{12}-J_{e}+E_{B}\end{array}\right).
\end{equation}

(8) $N=3$, $S_z=-\frac{1}{2}$. On the basis $\left\{ c_{1\uparrow}^{\dagger}c_{1\downarrow}^{\dagger}c_{2\downarrow}^{\dagger}\left|0\right\rangle, c_{2\uparrow}^{\dagger}c_{1\downarrow}^{\dagger}c_{2\downarrow}^{\dagger}\left|0\right\rangle \right\} $,
\begin{equation}
H_{8}=\left(\begin{array}{cc}
-2\mu_{1}-\mu_{2}+U_1+2U_{12}-J_{e}-E_{B} & -(t+J_{t1}+J_{t2})\\
-(t+J_{t1}+J_{t2}) & -2\mu_{2}-\mu_{1}+U_2+2U_{12}-J_{e}-E_{B}\end{array}\right).
\end{equation}
\end{widetext}

(9) $N=4$, $S_z=0$. On the basis $c_{1\uparrow}^{\dagger}c_{2\uparrow}^{\dagger}c_{1\downarrow}^{\dagger}c_{2\downarrow}^{\dagger}\left|0\right\rangle$,
\begin{equation}
H_{9}=-2\mu_{1}-2\mu_{2}+U_1+U_2+4U_{12}-2J_{e}.
\end{equation}

\renewcommand{\theequation}{B-\arabic{equation}}
\setcounter{equation}{0}
\section{Several extreme cases of the parameters}\label{extreme}

\begin{figure}[]
    \centering
    \includegraphics[width=5.5cm, angle=0]{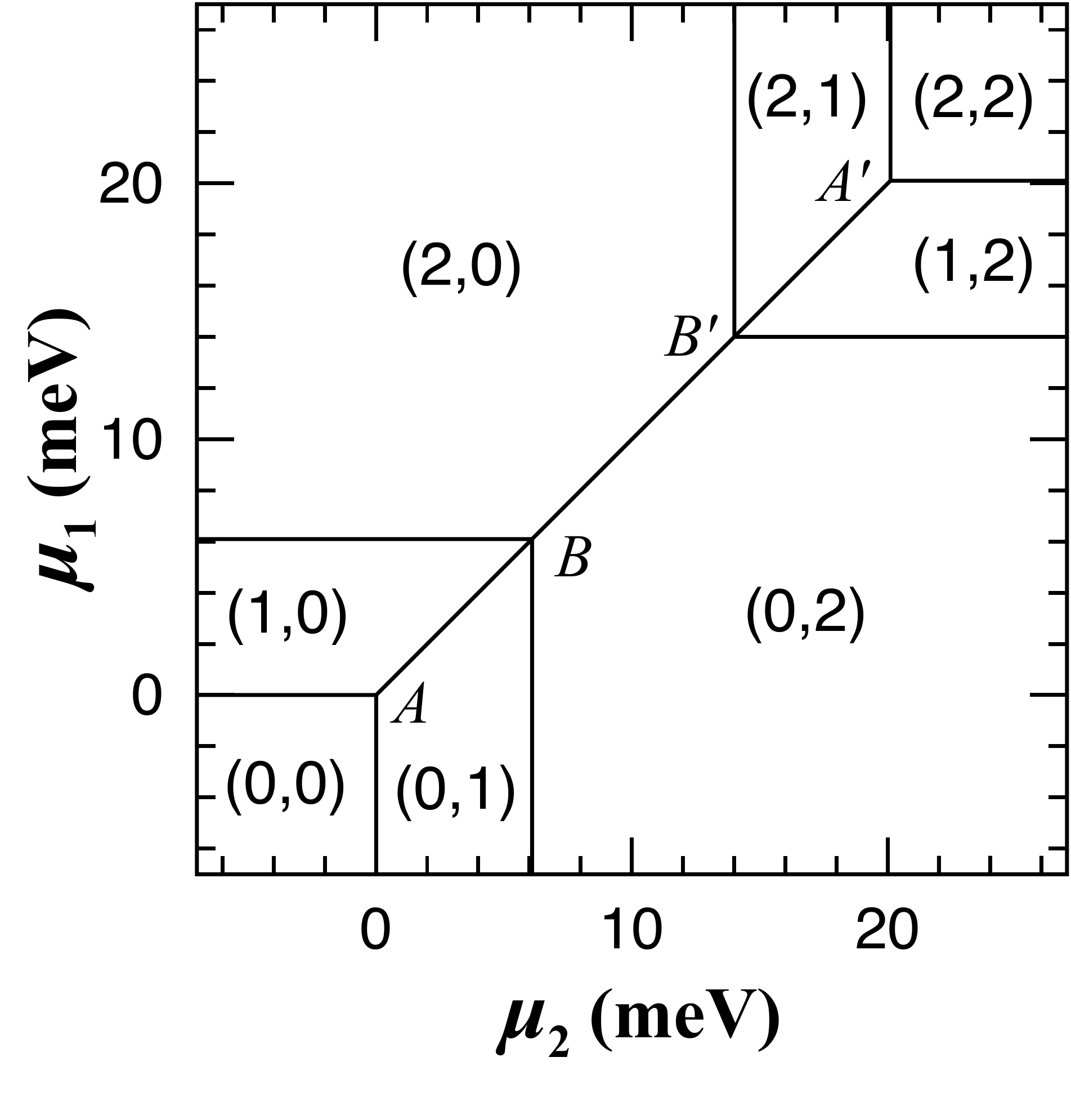}
    \caption{Charge stability diagram calculated at $U=6.1$ meV, $U_{12}=7$ meV, $t=J_e=J_p=J_t=0$.}
    \label{U12gtrU}
\end{figure}

\begin{figure}[]
    \centering
    (a)\includegraphics[width=5.5cm, angle=0]{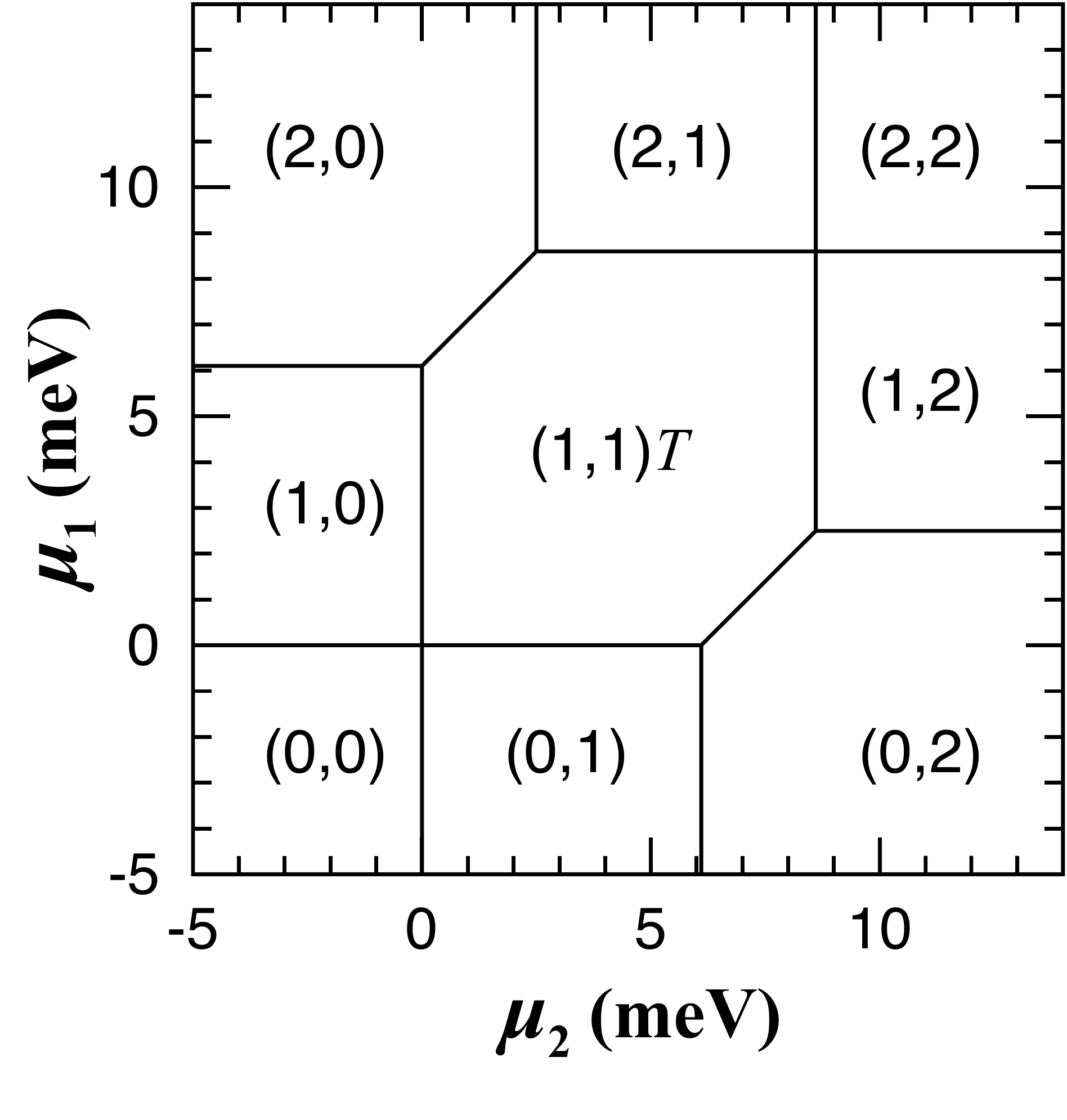}
    (b)\includegraphics[width=5.5cm, angle=0]{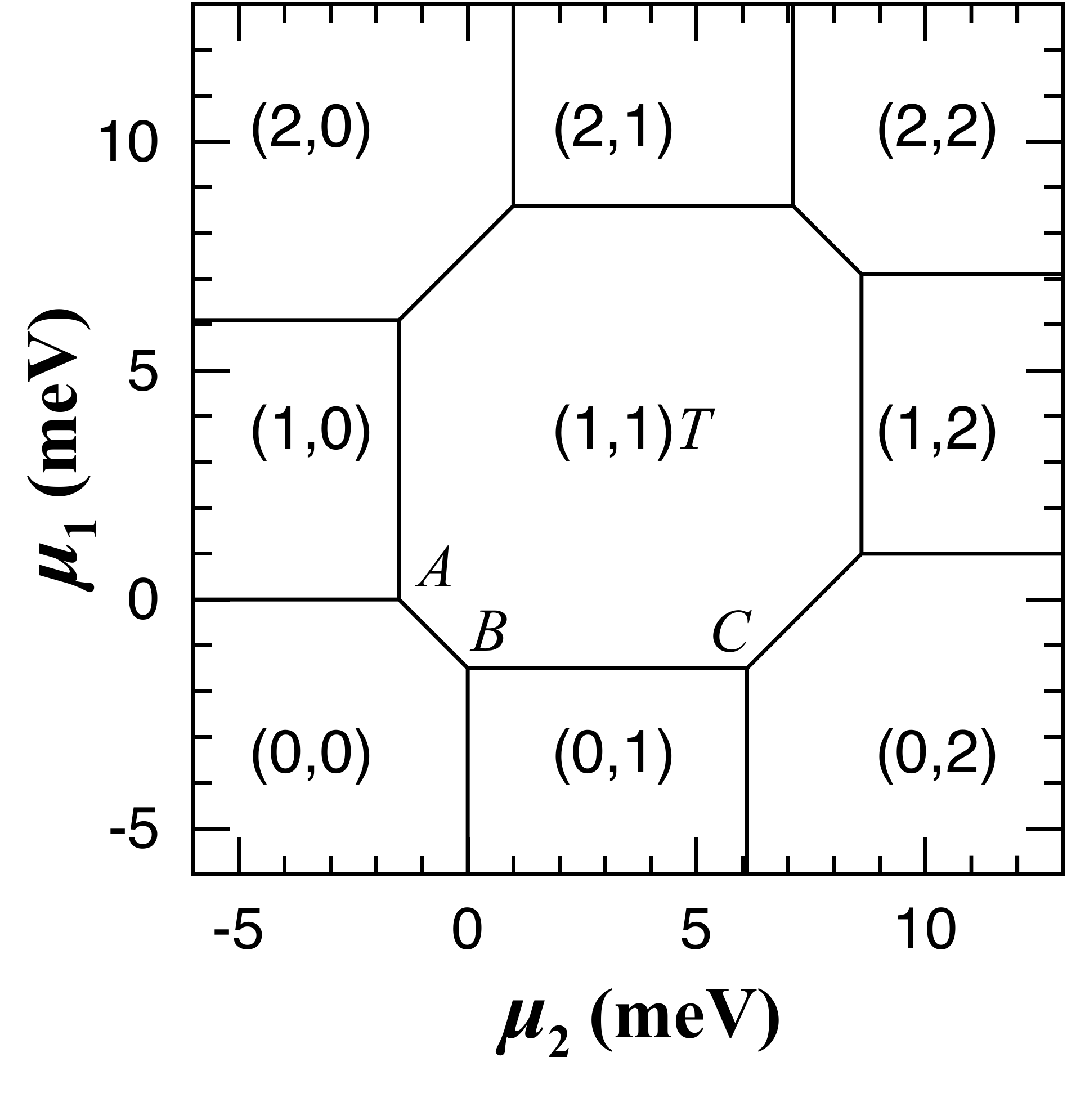}
    \caption{Charge stability diagram calculated at $U=6.1$ meV, $U_{12}=2.5$ meV, (a) $J_e=2.5~{\rm meV}=U_{12}$ (b) $J_e=4~{\rm meV}>U_{12}$. $t=J_p=J_t=0$.}
    \label{JegtrU12}
\end{figure}

In this section we present several extreme cases of parameters. As discussed above, $J_e<U_{12}<U$ in a typical physical system. However it remains interesting to study the cases with $U_{12}\ge U$ or $J_e\ge U_{12}$. First, for completeness of the understanding of the charge stability diagram, these parameter regimes need be explored. Indeed, as shall be seen below, the topology of the charge stability diagram changes. Second, although the systems we are currently studying impose special limitations of parameters, this in general need not be true. For example, in our study $J_e\ll t$. However, in a lattice problem the orthogonality of atomic orbitals means that $t=0$ while $J_e$ is still finite. Third, interesting physical phenomena happen in some of the parameter regimes, such as the $U_{12}>U$ case discussed in Ref.~\onlinecite{Wang.10}, which is at least of theoretical interest.

Fig.~\ref{U12gtrU} shows a typical case for $U_{12}\ge U$. The coordinates $(\mu_2,\mu_1)$ of the triple points are given by 
$A(0,0)$,
$B(U,U)$,
$B'(2U_{12},2U_{12})$,
$A'(U+2U_{12},U+2U_{12})$. The $(1,1)$ component is completely ruled out from the charge stability diagram since the system always favors either $(0,2)$ or $(2,0)$ state which have smaller Coulomb repulsion than the $(1,1)$ state. This charge stability diagram is not observed experimentally, which means that it is reasonable to set $U_{12}<U$ for currently implemented double quantum dot systems.

In Fig.~\ref{JegtrU12} we show the case with $J_e\ge U_{12}$ (but still $J_e<U$). Fig.~\ref{JegtrU12}(a) shows the $J_e=U_{12}$ result, which can be understood as the limit of Fig.~\ref{diagJe}.      Fig.~\ref{JegtrU12}(b) shows the result of    $U_{12}<J_e<U$. The topology fundamentally changes in that the (1,1) triplet state now shares a phase boundary directly with the (0,0) state. The coordinates $(\mu_2,\mu_1)$ of the triple points are
$A(U_{12}-J_e,0)$, $B(0,U_{12}-J_e)$, and $C(U, U_{12}-J_e)$, with others found by symmetry.
Interestingly the length of line segment $\overline{AB}$ is still $\sqrt{2}(J_e-U_{12})$ (which can be expressed as $\sqrt{2}\left|U_{12}-J_e\right|$ formally similar to that of Fig.~\ref{diagJe}).


\begin{thebibliography}{[00]}


\bibitem{Feynman.86}
R. P. Feynman, Int. J. Theor. Phys. {\bf 21}, 467 (1982); Found. Phys. {\bf 16}, 507 (1986); D. Deutsch, Proc. R. Soc. London, Ser. A {\bf 400}, 97 (1985)

\bibitem{Shor.94}
P. W. Shor, in \emph{Proceedings of the 35th Annual Symposium on the Foundations of Computer Science} (IEEE Press, Los Alamitos, CA, 1994), pp. 124133;
A. Ekert and R. Jozsa,  Rev. Mod. Phys. {\bf 68}, 733 (1996);
L. K. Grover, Phys. Rev. Lett. {\bf 79}, 325 (1997);
A. Steane, Rep. Prog. Phys. {\bf 61}, 117 (1998);
C. H. Bennett and D. P. DiVincenzo, Nature (London) {\bf 404}, 247 (2000).
 
\bibitem{Vandersypen.00}
L. M. K. Vandersypen, M. Steffen, G. Breyta, C. S. Yannoni, R. Cleve, and I. L. Chuang, Phys. Rev. Lett. {\bf 85}, 5452 (2000);
L. M. K. Vandersypen, M. Steffen, G. Breyta, C. S. Yannoni1, M. H. Sherwood, and I. L. Chuang, Nature (London) {\bf 414}, 883 (2001)

\bibitem{Sleator.95}
T. Sleator and H. Weinfurter, Phys. Rev. Lett. {\bf 74}, 4087 (1995); Q. A. Turchette, C. J. Hood, W. Lange, H. Mabuchi, and H.J. Kimble, \emph{ibid.} {\bf 75}, 4710 (1995).

\bibitem{Cirac.95}
J. I. Cirac and P. Zoller, Phys. Rev. Lett. {\bf 74}, 4091 (1995); C. Monroe, D. M. Meekhof, B. E. King, W. M. Itano, and D. J. Wineland, \emph{ibid.} {\bf 75}, 4714 (1995);
C. A. Sackett, D. Kielpinski, B. E. King, C. Langer, V. Meyer, C. J. Myatt, M. Rowe, Q. A. Turchette, W. M. Itano, D. J. Wineland, and C. Monroe, Nature (London) {\bf 404}, 256 (2000).


\bibitem{Loss.98}
D. Loss and D. P. DiVincenzo, Phys. Rev. A {\bf 57}, 120 (1998).
\bibitem{DiVincenzo.00}
D. P. DiVincenzo, D. Bacon, J. Kempe, G. Burkard, and K. B. Whaley, Nature {\bf 408}, 339 (2000).
\bibitem{Levy.02}
J. Levy, Phys. Rev. Lett.  {\bf 89}, 147902 (2002).
\bibitem{Koppens.06}
F. H. L. Koppens, C. Buizert, K. J. Tielrooij, I. T. Vink, K. C. Nowack, T. Meunier, L. P. Kouwenhoven, and L. M. K. Vandersypen, Nature {\bf 442}, 766 (2006).


\bibitem{Kane.98}
B. E. Kane, Nature {\bf 393}, 133 (1998).

\bibitem{Vrijen.00}
R. Vrijen, E. Yablonovitch, K. Wang, H. W. Jiang, A. Balandin, V. Roychowdhury, T. Mor, and D. DiVincenzo, Phys. Rev. A {\bf 62}, 012306 (2000).
\bibitem{Friesen.03}
M. Friesen, P. Rugheimer, D. E. Savage, M. G. Lagally, D. W. van der Weide, R. Joynt, and M. A. Eriksson, Phys. Rev. B {\bf 67}, 121301(R) (2003).
\bibitem{Taylor.05}
J. M. Taylor, H.-A. Engel, W. D\"{u}r, A. Yacoby, C. M. Marcus, P. Zoller, and M. D. Lukin, Nat. Phys. {\bf 1}, 177 (2005).


\bibitem{Kikkawa.98}
J. M. Kikkawa and D. D. Awschalom, Phys. Rev. Lett. {\bf 80}, 4313 (1998). 
\bibitem{Amasha.08}
S. Amasha, K. MacLean, Iuliana P. Radu, D. M. Zumb\"uhl, M. A. Kastner, M. P. Hanson, and A. C. Gossard, Phys. Rev. Lett. {\bf 100}, 046803 (2008). 
\bibitem{Koppens.08}
F. H. L. Koppens, K. C. Nowack, and L. M. K. Vandersypen, Phys. Rev. Lett. {\bf 100}, 236802 (2008). 
\bibitem{Barthel.10z}
C. Barthel, J. Medford, C. M. Marcus, M. P. Hanson, and A. C. Gossard, Phys. Rev. Lett. 105, 266808 (2010).
\bibitem{Bluhm.11}
H. Bluhm, S. Foletti, I. Neder, M. Rudner, D. Mahalu, V. Umansky, and A. Yacoby, Nat. Phys. {\bf 7}, 109 (2011). 

\bibitem{Tyryshkin.03}
A. M. Tyryshkin, S. A. Lyon, A. V. Astashkin, and A. M. Raitsimring, Phys. Rev. B {\bf 68}, 193207 (2003).
\bibitem{Morello.10}
A. Morello, J. J. Pla, F. A. Zwanenburg, K. W. Chan, K. Y. Tan, H. Huebl, M. M\"ott\"onen, C. D. Nugroho, C. Yang, J. A. van Donkelaar, A. D. C. Alves, D. N. Jamieson, C. C. Escott, L. C. L. Hollenberg, R. G. Clark, and A. S. Dzurak, Nature (London) {\bf 467}, 687 (2010).
\bibitem{Simmons.10}
C. B. Simmons, J. R. Prance, B. J. Van Bael, T. S. Koh, Z. Shi, D. E. Savage, M. G. Lagally, R. Joynt, M. Friesen, S. N. Coppersmith, and M. A. Eriksson, Phys. Rev. Lett. {\bf 106}, 156804 (2011). %e-print arXiv:1010.5828 (unpublished). 
\bibitem{Xiao.10a}
M. Xiao, M. G. House, and H. W. Jiang, Phys. Rev. Lett. {\bf 104}, 096801 (2010).


\bibitem{Petta.05}
J. R. Petta, A. C. Johnson, J. M. Taylor, E. A. Laird, A. Yacoby, M. D. Lukin, C. M. Marcus, M. P. Hanson, A. C. Gossard, Science {\bf 309}, 2180 (2005).
\bibitem{Foletti.09}
S. Foletti, H. Bluhm, D. Mahalu, V. Umansky, and A. Yacoby, Nat. Phys. {\bf 5}, 903 (2009).
\bibitem{Laird.10}
E. A. Laird, J. M. Taylor, D. P. DiVincenzo, C. M. Marcus, M. P. Hanson, and A. C. Gossard, Phys. Rev. B {\bf 82}, 075403 (2010).
\bibitem{vanWeperen.11}
I. van Weperen, B. D. Armstrong, E. A. Laird, J. Medford, C. M. Marcus, M. P. Hanson, and A. C. Gossard, Phys. Rev. Lett. {\bf 107}, 030506 (2011).

\bibitem{Nordberg.09}
E. P. Nordberg, G. A. Ten Eyck, H. L. Stalford, R. P. Muller, R. W. Young, K. Eng, L. A. Tracy, K. D. Childs, J. R. Wendt, R. K. Grubbs, J. Stevens, M. P. Lilly, M. A. Eriksson, and M. S. Carroll, Phys. Rev. B {\bf 80}, 115331 (2009).

\bibitem{dassarma.05}
S. Das Sarma, R. de Sousa, X. Hu, and B. Koiller, Sol. St. Comm. {\bf 133}, 737 (2005).
\bibitem{Culcer.10}
D. Culcer, \L. Cywi\'nski, Q. Li, X. Hu, and S. Das Sarma, Phys. Rev. B {\bf 80}, 205302 (2009); {\bf 82}, 155312 (2010).
\bibitem{Goswami.07}
S. Goswami, K. A. Slinker, M. Friesen, L. M. McGuire, J. L. Truitt, C. Tahan, L. J. Klein, J. O. Chu, P. M. Mooney, D. W. van der Weide, R. Joynt, S. N. Coppersmith, and M. A. Eriksson, Nat. Phys. {\bf 3}, 41 (2007).
\bibitem{Borselli.11}
M. G. Borselli, R. S. Ross, A. A. Kiselev, E. T. Croke, K. S. Holabird, P. W. Deelman, L. D. Warren, I. Alvarado-Rodriguez, I. Milosavljevic, F. C. Ku, W. S. Wong, A. E. Schmitz, M. Sokolich, M. F. Gyure, and A. T. Hunter, Appl. Phys. Lett. {\bf 98}, 123118 (2011).


\bibitem{Simmons.09}
C. B. Simmons, M. Thalakulam, B. M. Rosemeyer, B. J. Van Bael, E. K. Sackmann, D. E. Savage, M. G. Lagally, R. Joynt, M. Friesen, S. N. Coppersmith, and M. A. Eriksson, Nano Lett. {\bf 9} 3234 (2009).
\bibitem{Lai.10}
N. S. Lai, W. H. Lim, C. H. Yang, F. A. Zwanenburg, W. A. Coish, F. Qassemi, A. Morello, and A. S. Dzurak, e-print arXiv:1012.1410 (unpublished).


\bibitem{Livermore.96}
C. Livermore, C. H. Crouch, R. M. Westervelt, K. L. Campman, A. C. Gossard, Science {\bf 274}, 1332 (1996).


\bibitem{Elzerman.03}
J. M. Elzerman, R. Hanson, J. S. Greidanus, L. H. Willems van Beveren, S. De Franceschi, L. M. K. Vandersypen, S. Tarucha, and L. P. Kouwenhoven, Phys. Rev. B {\bf 67}, 161308(R) (2003).



\bibitem{Hofmann.95}
F. Hofmann, T. Heinzel, D.A. Wharam, J.P. Kotthaus, G. B\"ohm, W. Klein, G. Tr\"ankle, and G. Weimann, Phys. Rev. B {\bf 51}, 13872 (1995).
\bibitem{DiCarlo.04}
L. DiCarlo, H. J. Lynch, A. C. Johnson, L. I. Childress, K. Crockett, C. M. Marcus, M. P. Hanson, and A. C. Gossard, Phys. Rev. Lett. {\bf 92}, 226801 (2004).


\bibitem{Scarola.04}
V. W. Scarola, K. Park, and S. Das Sarma, Phys. Rev. Lett. {\bf 93}, 120503 (2004); V. W. Scarola and S. Das Sarma, Phys. Rev. A {\bf 71}, 032340 (2005).

\bibitem{LiQZ.10}
Q. Li, \L. Cywi\'nski, D. Culcer, X. Hu, and S. Das Sarma, Phys. Rev. B {\bf 81}, 085313 (2010).

\bibitem{Taylor.07}
J. M. Taylor, J. R. Petta, A. C. Johnson, A. Yacoby, C. M. Marcus, and M. D. Lukin, Phys. Rev. B {\bf 76}, 035315 (2007)


\bibitem{Petta.08}
J. R. Petta, J. M. Taylor, A. C. Johnson, A. Yacoby, M. D. Lukin, C. M. Marcus, M. P. Hanson, and A. C. Gossard,
Phys. Rev. Lett. {\bf 100}, 067601 (2008).

\bibitem{Reilly.08}
D. J. Reilly, J. M. Taylor, J. R. Petta, C. M. Marcus, M. P. Hanson, and A. C. Gossard,
Science {\bf 321}, 817 (2008).

\bibitem{Reilly.10}
D. J. Reilly, J. M. Taylor, J. R. Petta, C. M. Marcus, M. P. Hanson, and A. C. Gossard,
Phys. Rev. Lett. {\bf 104}, 236802 (2010).

\bibitem{Barthel.09}
C. Barthel, D. J. Reilly, C. M. Marcus, M. P. Hanson, and A. C. Gossard,
Phys. Rev. Lett. {\bf 103}, 160503 (2009).

\bibitem{Barthel.10a}
C. Barthel, M. Kj\ae{}rgaard, J. Medford, M. Stopa, C. M. Marcus, M. P. Hanson, and A. C. Gossard,
Phys. Rev. B {\bf 81} 161308(R), (2010).

\bibitem{Beenakker.91}
C. W. J. Beenakker, Phys. Rev. B {\bf 44}, 1646 (1991)
\bibitem{Wiel.03}
W. G. van der Wiel, S. De Franceschi, J. M. Elzerman, T. Fujisawa, S. Tarucha, L. P. Kouwenhoven, Rev. Mod. Phys. {\bf 75}, 1 (2003).
\bibitem{Hanson.07}
R. Hanson, L. P. Kouwenhoven, J. R. Petta, S. Tarucha, L. M. K. Vandersypen, Rev. Mod. Phys. {\bf 79}, 1217 (2007).

\bibitem{Schroer.07}
D. Schr\"{o}er, A. D. Greentree, L. Gaudreau, K. Eberl, L. C. L. Hollenberg, J. P. Kotthaus, and S. Ludwig, Phys. Rev. B {\bf 76}, 075306 (2007).


\bibitem{Yang.11}
S. Yang, X. Wang, and S. Das Sarma, Phys. Rev. B {\bf 83}, 161301(R) (2011).
\bibitem{dassarma.Si}
S. Das Sarma, X. Wang, and S. Yang, Phys. Rev. B {\bf 83}, 235314 (2011).


\bibitem{Petta.04}
J. R. Petta, A. C. Johnson, C. M. Marcus, M. P. Hanson, and A. C. Gossard,  Phys. Rev. Lett. {\bf 93}, 186802 (2004)
\bibitem{Hatano.05}
T. Hatano, M. Stopa, and S. Tarucha, Science {\bf 309}, 268 (2005).

\bibitem{Huttel.05}
A. K. H\"{u}ttel, S. Ludwig, H. Lorenz, K. Eberl, and J. P. Kotthaus, Phys. Rev. B {\bf 72}, 081310 (2005).

\bibitem{Pioro.05}
M. Pioro-Ladri\`{e}re, M. R. Abolfath, P. Zawadzki, J. Lapointe, S. A. Studenikin, A. S. Sachrajda, and P. Hawrylak, Phys. Rev. B {\bf 72}, 125307 (2005).

\bibitem{ZhangLX.06}
L.-X. Zhang, D. V. Melnikov, and J.-P. Leburton, Phys. Rev. B {\bf 74}, 205306 (2006).

\bibitem{Stafford.94}
C. A. Stafford and S. Das Sarma, Phys. Rev. Lett. {\bf 72}, 3590 (1994).
\bibitem{Kotlyar.98}
R. Kotlyar, C. A. Stafford, and S. Das Sarma, Phys. Rev. B {\bf 58}, R1746 (1998); {\bf 58}, 3989 (1998); C. A. Stafford, R. Kotlyar, and S. Das Sarma, \emph{ibid.} {\bf 58}, 7091 (1998).
\bibitem{Jefferson.96}
J. H. Jefferson and W. H\"ausler, Phys. Rev. B {\bf 54}, 4936 (1996).

\bibitem{Gaudreau.06}
L. Gaudreau, S. A. Studenikin, A. S. Sachrajda, P. Zawadzki, A. Kam, J. Lapointe, M. Korkusinski, and P. Hawrylak, Phys. Rev. Lett. {\bf 97}, 036807 (2006).
\bibitem{Korkusinski.07}
M. Korkusinski, I. P. Gimenez, P. Hawrylak, L. Gaudreau, S. A. Studenikin, and A. S. Sachrajda, Phys. Rev. B {\bf 75}, 115301 (2007).
%%%%%%newrefs
%nuclearspin
%\bibitem{Reilly.08}
%\bibitem{Reilly.10}{Taylor.07}
\bibitem{Witzel.06}
W. M. Witzel and S. Das Sarma, Phys. Rev. B {\bf 74}, 035322 (2006)
\bibitem{Cywinski.09}
\L. Cywi\'nski, W. M. Witzel, and S. Das Sarma, Phys. Rev. Lett. {\bf 102}, 057601 (2009); Phys. Rev. B {\bf 79}, 245314 (2009)
%%impurity
\bibitem{Gimenez.09}
I. P. Gimenez, C.-Y. Hsieh, M. Korkusinski, and P. Hawrylak, Phys. Rev. B {\bf 79}, 205311 (2009)
\bibitem{Nguyen.11}
N. T. T. Nguyen and S. Das Sarma, Phys. Rev. B {\bf 83}, 235322 (2011)
%phonon
\bibitem{Storcz.05}
M. J. Storcz, U. Hartmann, S. Kohler, and F. K. Wilhelm, Phys. Rev. B {\bf 72}, 235321 (2005)
\bibitem{Stavrou.05}
V. N. Stavrou and X. Hu, Phys. Rev. B {\bf 72}, 075362 (2005)
\bibitem{Stano.06}
P. Stano and J. Fabian, Phys. Rev. Lett. {\bf 96}, 186602 (2006)
\bibitem{Harbusch.10}
D. Harbusch, D. Taubert, H. P. Tranitz, W. Wegscheider, and S. Ludwig, Phys. Rev. Lett. {\bf 104}, 196801 (2010)
\bibitem{Hu.11}
X. Hu, Phys. Rev. B {\bf 83}, 165322 (2011)
%%%%%%%%%

\bibitem{Artacho.91}
E. Artacho and L. Mil\'ans del Bosch, Phys. Rev. A {\bf 43}, 5770 (1991).

\bibitem{Hu.00}
X. Hu and S. Das Sarma, Phys. Rev. A {\bf 61}, 062301 (2000). 


\bibitem{Slater.36}
J. C. Slater, Phys. Rev. {\bf 49}, 537 (1936). 

\bibitem{Anderson.59}
P. W. Anderson, Phys. Rev. {\bf 115}, 2 (1959).
\bibitem{Kanamori.63}
J. Kanamori, J. Phys. Chem. Solids {\bf 10}, 87 (1959); Prog. Theor. Phys. {\bf 30}, 275 (1963). 

\bibitem{Hirsch.92}
J. E. Hirsch, Physica C {\bf 199}, 305 (1992);
J. E. Hirsch and F. Marsiglio, Phys. Rev. B {\bf 62}, 15131 (2000). 

\bibitem{Hubbard.63}
J. Hubbard, Proc. R. Soc. London, Ser. A {\bf 276}, 238 (1963); 
{\bf 277}, 237 (1964);
{\bf 281}, 401 (1964).
\bibitem{Zaanen.96}
J. Zaanen and M. Ole\'s, Ann. Phys. (Leipzig) {\bf 5}, 224 (1996)
\bibitem{Imada.98}
M. Imada, A. Fujimori, Y. Tokura, Rev. Mod. Phys. {\bf 70}, 1039 (1998). 
\bibitem{Liebsch.03}
A. Liebsch, Phys. Rev. Lett. 91, 226401 (2003);
A. Koga, N. Kawakami, T. M. Rice, and M. Sigrist, Phys. Rev. Lett. {\bf 92}, 216402 (2004);
P. Werner and A. J. Millis, Phys. Rev. B {\bf 74}, 155107 (2006) 

\bibitem{Cottet.11}
A. Cottet, C. Mora, and T. Kontos, Phys. Rev. B {\bf 83}, 121311(R) (2011).

\bibitem{Burkard.99}
G. Burkard, D. Loss, and D. P. DiVincenzo, Phys. Rev. B {\bf 59}, 2070 (1999).
\bibitem{Hu.01}
X. Hu and S. Das Sarma, Phys. Rev. A {\bf 64}, 042312 (2001).
\bibitem{Sousa.01}
R. de Sousa, X. Hu, and S. Das Sarma, Phys. Rev. A {\bf 64}, 042307 (2001).

\bibitem{Gimenez.07}
I. Puerto Gimenez, M. Korkusinski, and P. Hawrylak, Phys. Rev. B {\bf 76}, 075336 (2007).
\bibitem{Nielsen.10}
E. Nielsen, R. W. Young, R. P. Muller, and M. S. Carroll, Phys. Rev. B {\bf 82}, 075319 (2010)
\bibitem{Nielsen.11}
E. Nielsen and R. P. Muller, e-print arXiv:1006.2735 (unpublished)

\bibitem{Helle.05}
M. Helle, A. Harju, and R. M. Nieminen, Phys. Rev. B {\bf 72}, 205329 (2005)
\bibitem{Pedersen.07}
J. Pedersen, C. Flindt, N. A. Mortensen, and A.-P. Jauho, Phys. Rev. B {\bf 76}, 125323 (2007)


\bibitem{Pedersen.10}
J. G. Pedersen, C. Flindt, A.-P. Jauho, and N. A. Mortensen, Phys. Rev. B {\bf 81}, 193406 (2010). 

\bibitem{Hu.06}
X. Hu and S. Das Sarma, Phys. Rev. Lett. 96, 100501 (2006)

\bibitem{Stopa.08}
M. Stopa, and C. M. Marcus, Nano Lett., {\bf 8}, 1778 (2008).

\bibitem{Witzel.05}
W. M. Witzel, R. de Sousa, and S. Das Sarma, Phys. Rev. B {\bf 72}, 161306(R) (2005)
\bibitem{Witzel.08}
W. M. Witzel and S. Das Sarma, Phys. Rev. B {\bf 77}, 165319 (2008) 

\bibitem{Witzel.10}
W. M. Witzel, M. S. Carroll, A. Morello, \L. Cywi\'nski, and S. Das Sarma, Phys. Rev. Lett. {\bf 105}, 187602 (2010)

\bibitem{Wang.10}
X. Wang and A. J. Millis, Phys. Rev. B {\bf 81}, 045106 (2010). 


\end{thebibliography}
\end{document}